\documentclass{llncs}

\def\todoFlag{ON} 
\def\longFlag{OFF} 

\newif\ifcomments
\commentstrue

\usepackage{isabelle,isabellesym}
\usepackage{amsfonts}
\usepackage{latexsym}
\usepackage{amssymb}
\usepackage{amsmath}
\usepackage{alltt}
\usepackage{stmaryrd}

\usepackage{url}
\usepackage{color}
\usepackage{fancyvrb}

\definecolor{darkgreen}{rgb}{0.0,0.5,0.0}
\ifcomments
  \newcommand{\flp}[1]{\par\noindent\textcolor{blue}{\sf FLP: #1}}
  \newcommand{\sm}[1]{\par\noindent\textcolor{darkgreen}{\sf SM: #1}}
  \newcommand{\jrh}[1]{\par\noindent\textcolor{red}{\sf JRH: #1}}
  \newcommand{\FLP}[1]{\marginpar{\footnotesize\textcolor{blue}{\sf FLP: #1}}}
  \newcommand{\SM}[1]{\marginpar{\footnotesize\textcolor{darkgreen}{\sf SM: #1}}}
  \newcommand{\JRH}[1]{\marginpar{\footnotesize\textcolor{red}{\sf JRH: #1}}}
\else
  \newcommand{\flp}[1]{}
  \newcommand{\sm}[1]{}
  \newcommand{\jrh}[1]{}
  \newcommand{\FLP}[1]{}
  \newcommand{\SM}[1]{}
  \newcommand{\JRH}[1]{}
\fi

\usepackage{ifthen} 
\newcommand{\nc}{\newcommand}

\newcommand{\den}[1]{[\![#1]\!]}
\newcommand{\denn}[1]{[\![\![#1]\!]\!]}

\nc{\ordSus}{\lesssim} 
\nc{\ordap}{\sqsubseteq} 
\nc{\rw}{\to} 
\nc{\tor}{\to} 
\nc{\crwlto}{\rightarrowtriangle}  
\nc{\clto}{\crwlto}                
\nc{\conscrwl}{\vdash_{\crwl}}                
\nc{\clp}{{\cal P} \vdash_{CRWL^{+}}} 
\nc{\cldt}{{\cal P} \vdash_{CRWL^{d}}} 
\nc{\dend}[1]{\den{#1}^d} 
\nc{\dg}[1]{\den{#1}_{CRWL}}
\nc{\cl}{{\cal P} \vdash_{CRWL}}
\nc{\clho}{{\cal P} \vdash_{HOCRWL}}
\nc{\gl}{{\cal P} \vdash_{CRWL_{let}}}
\nc{\dgl}[1]{\den{#1}_{CRWL_{let}}}
\nc{\dcl}[1]{\dgl{#1}}
\nc{\ddgl}[1]{\denn{#1}_{CRWL_{let}}}
\nc{\ddcl}[1]{\ddgl{#1}}

\nc{\var}{{\cal V}}
\nc{\ra}{\tor}
\nc{\leqhyp}{\Subset}
\nc{\tot}[1]{{#1}^\tau}
\nc{\tlr}[1]{\widehat{#1}} 
\nc{\jn}{\Join} 
\nc{\tr}{\underline{\mbox{\textbf{t}}}}
\nc{\slt}{\hookleftarrow} 
\nc{\clt}{\hookrightarrow} 
\nc{\con}{{\cal C}}  
\nc{\cnn}[1]{\con[#1]}  
\nc{\cnnp}[1]{\con'[#1]}  
\nc{\nat}{{\mathbb N}}
\nc{\prog}{{\cal P}} 
\nc{\progh}{\hat{\prog}} 
\nc{\progm}{{\cal M}} 
\nc{\progr}{{\cal R}} 
\nc{\ps}{\vdash}    
\nc{\pps}{\vdash_{\pi CRWL}}    
\nc{\pcrwl}{{\it $\pi$CRWL}} 
\nc{\crwl}{\emph{CRWL}}
\nc{\denp}[1]{\den{#1}^{pl}} 
\nc{\dens}[1]{\den{#1}} 
\nc{\denr}[1]{\den{#1}^{rw}} 
\nc{\pst}[1]{pST(#1)} 
\nc{\icsus}{CSubst_\perp^?}
\nc{\ppop}{\rightarrowtail} 
\nc{\pordap}{\ordap_{\pi}} 
\nc{\dsord}{\unlhd} 
\nc{\cltop}{\clto} 
\nc{\dsordp}{\dsord_{M}} 
\nc{\pl}{\pi } 
\nc{\cjto}{{\cal A}} 
\nc{\cjtoD}{{\cal S}} 
\nc{\vran}{vran}
\nc{\trs}{{TRS}} 
\nc{\ctrs}{CS} 
\nc{\trss}{\trs's} 
\nc{\ctrss}{\ctrs's} 
\nc{\crule}[1]{\textsc{#1}} 
\nc{\isaelem}[1]{\texttt{#1}}

\nc{\todo}[1]{\ifthenelse{\equal{\todoFlag}{ON}}{~\\\textcolor[rgb]{1.00,0.00,0.00}{{\fbox{\begin{minipage}{.95\textwidth}{#1}\end{minipage}}}}}{}
} 
\nc{\longText}[2]{\ifthenelse{\equal{\longFlag}{ON}}{{#1}}{{#2}}} 
\nc{\regla}[4]{
\textbf{({#1})} $\ $ $\begin{array}{c}
{#3}\\
\hline
{#2}
\end{array}$
\quad {#4}
}
\newenvironment{isacode}
{\begin{list}{}{
\setlength{\leftmargin}{4pt}
\setlength{\rightmargin}{0pt}
\setlength{\listparindent}{0pt}
\raggedright
\setlength{\itemsep}{0pt}
\setlength{\parsep}{0pt}
\normalfont\ttfamily 
}%
 \item[]}
{\end{list}}

\makeatletter
\newcounter{abr@ctr}
\newcommand{\abr@c}{\c@abr@ctr\advance\c@abr@ctr\@ne}

\ifx\documentclass\undefined
\else
  \DeclareSymbolFont{tlaitalics}{\encodingdefault}{cmr}{m}{it}
  \let\itfam\symtlaitalics
\fi

\newcommand{\noTeXmath}{%
\c@abr@ctr=\itfam
\multiply\c@abr@ctr"100\relax
\advance\c@abr@ctr "7061\relax
\mathcode`a=\abr@c\mathcode`b=\abr@c\mathcode`c=\abr@c\mathcode`d=\abr@c
\mathcode`e=\abr@c\mathcode`f=\abr@c\mathcode`g=\abr@c\mathcode`h=\abr@c
\mathcode`i=\abr@c\mathcode`j=\abr@c\mathcode`k=\abr@c\mathcode`l=\abr@c
\mathcode`m=\abr@c\mathcode`n=\abr@c\mathcode`o=\abr@c\mathcode`p=\abr@c
\mathcode`q=\abr@c\mathcode`r=\abr@c\mathcode`s=\abr@c\mathcode`t=\abr@c
\mathcode`u=\abr@c\mathcode`v=\abr@c\mathcode`w=\abr@c\mathcode`x=\abr@c
\mathcode`y=\abr@c\mathcode`z=\abr@c\c@abr@ctr=\itfam
\multiply\c@abr@ctr"100\relax
\advance\c@abr@ctr "7041\relax
\mathcode`A=\abr@c\mathcode`B=\abr@c\mathcode`C=\abr@c\mathcode`D=\abr@c
\mathcode`E=\abr@c\mathcode`F=\abr@c\mathcode`G=\abr@c\mathcode`H=\abr@c
\mathcode`I=\abr@c\mathcode`J=\abr@c\mathcode`K=\abr@c\mathcode`L=\abr@c
\mathcode`M=\abr@c\mathcode`N=\abr@c\mathcode`O=\abr@c\mathcode`P=\abr@c
\mathcode`Q=\abr@c\mathcode`R=\abr@c\mathcode`S=\abr@c\mathcode`T=\abr@c
\mathcode`U=\abr@c\mathcode`V=\abr@c\mathcode`W=\abr@c\mathcode`X=\abr@c
\mathcode`Y=\abr@c\mathcode`Z=\abr@c}

\makeatother
\noTeXmath

\title{A Formalization of the Semantics of Functional-Logic Programming in Isabelle\thanks{This work has been partially supported by the Spanish projects TIN2005-09207-C03-03 (MERIT-FORMS-UCM), S-0505/TIC/0407 (PROMESAS-CAM) and TIN2008-06622-C03-01/TIN (FAST-STAMP).}}

\author{Francisco J. L\'opez-Fraguas\inst{1} \and Stephan Merz\inst{2} \and Juan Rodr\'iguez-Hortal\'a\inst{1} }

\institute{Departamento de Sistemas Inform\'aticos y Computaci\'on\\Universidad Complutense de Madrid, Spain \\
\email{fraguas@sip.ucm.es, juanrh@fdi.ucm.es} \and
INRIA Nancy \& LORIA \\ \email{Stephan.Merz@loria.fr}
}

\begin{document}

\maketitle

\begin{abstract}
  Modern functional-logic programming languages like Toy or Curry
  feature non-strict non-deterministic functions that behave under
  call-time choice semantics. A standard formulation for this
  semantics is the \crwl\ logic, that specifies a proof calculus for
  computing the set of possible results for each expression. In this
  paper we present a formalization of that calculus in the
  Isabelle/HOL proof assistant. We have proved some basic properties
  of \crwl: closedness under c-substitutions, polarity and
  compositionality. We also discuss some insights that have been
  gained, such as the fact that left linearity of program rules is not
  needed for any of these results to hold.
\end{abstract}

\section{Introduction}\label{intro}
\longText{%
  A key aspect of research on any programming language paradigm is
  providing precise theoretical foundations, usually in the form of
  some mathematical theory for the (denotational, operational)
  semantics of the language. This is specially true in the case of
  \emph{declarative} languages, that frequently are designed with an
  underlying corresponding theory or formalism in mind (e.g., Horn
  logic for logic programming, $\lambda$-calculus or term rewriting
  for functional programming). Fully formalizing the (meta)theory of a
  programming language or paradigm is a further step that can be done
  in the development of its foundations. There is an increasing number
  of researchers (see e.g. \cite{DBLP:conf/tphol/AydemirBFFPSVWWZ05})
  sharing the conviction that the combination
  \emph{formalization+mechanized theorem proving} must (and will) play
  a prominent role in the next future of programming languages
  research and technology. Just a couple of reasons: formalizations
  help to clarify overlooked aspects or overcome common
  misunderstandings, to discover some pitfalls, or even to provide new
  insights; moreover, formalized
  metatheories 
  are a direct vehicle to mechanized reasoning about programs, and
  therefore a support for software development tools like certifying
  compilers or certified program transformations.
}{%
  Fully formalizing the (meta)theory of a programming language can be
  beneficial for developing its foundations. There is an increasing
  number of researchers (see e.g.
  \cite{DBLP:conf/tphol/AydemirBFFPSVWWZ05}) sharing the conviction
  that the combination \emph{formalization+mechanized theorem proving}
  must (and will) play a prominent role in 
  programming languages research and technology. In particular,
  formalizations help to clarify overlooked aspects, to discover
  pitfalls, and even to provide new insights; moreover, formalized
  metatheories 
  lead to mechanized reasoning about programs, giving reliable support
  to tools like certifying compilers or certified program
  transformations.
}

\longText{%
  In this paper we address the problem of formalizing the semantics of
  functional logic programming (FLP), a well established paradigm
  aiming at integrating the best features of logic and functional
  languages (see \cite{Hanus07ICLP} for a recent survey on FLP).
}{%
  In this paper we formalize the semantics of functional logic
  programming (FLP), a well established paradigm
  (see~\cite{Hanus07ICLP}) integrating features of logic and
  functional languages.
}
In modern FLP languages such as Curry~\cite{Han06curry} or
Toy~\cite{LS99} programs are constructor based rewrite systems that
may be non-terminating and non-confluent. Semantically this leads to
the presence of non-strict and non-deterministic functions. The
semantics adopted for non-determinism is \emph{call-time
  choice}~\cite{hussmann93,GHLR99}, informally meaning that in any
reduction, all descendants of a given subexpression must share the
same value. The semantic framework \crwl%
\footnote{\crwl{} stands for ``Constructor-based ReWriting Logic''.}
was proposed in~\cite{GHLR96,GHLR99} to accomodate this view of
non-determinism, and is nowadays considered the standard semantics of
FLP. For the purpose of this paper, the most relevant aspect of
\crwl{} is a proof calculus devised to prove reduction statements of
the form $\prog \vdash e \crwlto t$, meaning that $t$ is a possible
(partial) value to which $e$ can be reduced using the program $\prog$.


We have chosen Isabelle/HOL as concrete logical framework for our
formalization. Using such a broadly used system is not only easier,
but also more flexible and stable than developing language specific
tools like has been done, e.g., for logic programming~\cite{Stark98}
or functional programming~\cite{MolEP01}.

The remainder of the paper is organized as follows:
Sect.~\ref{preliminaries} contains some preliminaries about the
\crwl{} framework \longText{and the Isabelle system}{},
Sect.~\ref{formalizing} presents the Isabelle theories developed to
formalize \crwl{}, and Sect.~\ref{properties} gives the mechanized
proofs of some important properties of \crwl{}. Finally,
Sect.~\ref{conclusions} summarizes some conclusions and points to
future work.

\longText{}{%
  An extended version of this paper can be found at
  \url{http://gpd.sip.ucm.es/juanrh/pubs/isabell-crwl-report.pdf}.
}
The Isabelle code underlying the results presented here is available
at \url{https://gpd.sip.ucm.es/trac/gpd/wiki/GpdSystems/IsabelleCrwl}.


\section{Preliminaries}\label{preliminaries}
\subsection{Constructor-based term rewrite systems}
\label{trs}

We consider a first-order signature
$\Sigma = CS\cup FS$, where $CS$ and $FS$ are two disjoint sets of \emph{constructor}
and defined \emph{function} symbols respectively, each with associated arity. We
write $CS^n$ ($FS^n$ resp.) for the set of constructor (function) symbols of
arity $n$.
%
The set $Exp$ of \emph{expressions} is inductively defined as
\[
  Exp \ni e::= X \mid h(e_1,\ldots,e_n),
\]
where $X\in\var$, $h\in CS^n\cup FS^n$ and $e_1,\ldots,e_n\in Exp$.
The set $CTerm$ of \emph{constructed terms} (or \emph{c-terms}) is
defined like $Exp$, but with $h$ restricted to $CS^n$ (so
$CTerm \subseteq Exp$). The intended meaning is that $Exp$ stands for
evaluable expressions, i.e., expressions that can contain function
symbols, while $CTerm$ stands for data terms representing values. We
will write $e,e',\ldots$ for expressions and $t,s,\ldots$ for c-terms.
The set of variables occurring in an expression $e$ will be denoted as
$var(e)$.
We will frequently use \emph{one-hole  contexts}, defined as
\[
  Cntxt \ni {\cal C} ::= [\ ] \mid h(e_1,\ldots,{\cal C},\ldots,e_n)
\]
for $h\in CS^n\cup FS^n$. The application of a context ${\cal C}$ to
an expression $e$, written ${\cal C}[e]$, is defined inductively by
\[
  [\ ][e] = e \quad\text{and}\quad
  h(e_1,\ldots,{\cal C},\ldots,e_n)[e] = h(e_1,\ldots,{\cal C}[e],\ldots,e_n).
\]

The set \emph{Subst} of \emph{substitutions} consists of finite
mappings $\theta:\var \longrightarrow Exp$ (i.e., mappings such that
$\theta(X) \neq X$ only for finitely many $X\in\var$), which extend
naturally to $\theta:Exp \longrightarrow
Exp$. 
We write $e\theta$ for the application of $\theta$ to $e$, and
$\theta\theta'$ for the composition of substitutions, defined by
$X(\theta\theta') = (X\theta)\theta'$. The domain 
of $\theta$ 
is defined as $dom(\theta) = \{X\in \var \mid X\theta \neq
X\}$. 
In most cases we will use \emph{c-substitutions}
$\theta \in \emph{CSubst}$, for which $X\theta \in CTerm$ for all $X\in dom(\theta)$.
%


A \emph{\crwl{}-program} (or simply a \emph{program})
is a set of rewrite rules of the form $f(\overline{t})\to e$ where
$f\in FS^n$, $e\in Exp$ and $\overline{t}$ is a linear $n$-tuple of
c-terms, where linearity means that each variable occurs only once in
$\overline{t}$. Notice that we allow $e$ to contain \emph{extra
  variables}, i.e., variables not occurring in $\overline{t}$.
\crwl{}-programs often
allow also conditions in the program rules. However, \crwl{}-programs
with conditions can be transformed into equivalent programs without conditions,
therefore we consider only unconditional rules.

\subsection{The \crwl{} framework}


In order to accomodate non-strictness at the semantic level, we
enlarge $\Sigma$ with a new constant constructor symbol $\perp$. The
sets $Exp_\perp$, $CTerm_\perp$, $Subst_\perp$, $CSubst_\perp$ of
partial expressions, etc., are defined naturally. Notice that $\perp$
does not appear in programs. Partial expressions are ordered by the
\emph{approximation} ordering $\sqsubseteq$ defined as the least
partial ordering satisfying
\[
  \mathop{\perp} \sqsubseteq e \quad\text{and}\quad
  e \sqsubseteq e' \Rightarrow {\cal C}[e] \sqsubseteq {\cal C}[e']
  \text{ for all } e,e' \in Exp_\perp, {\cal C} \in Cntxt
\]
This partial ordering can be extended to substitutions: given
$\theta,\sigma\in Subst_\bot$ we say $\theta\sqsubseteq\sigma$ if
$X\theta\sqsubseteq X\sigma$ for all $X\in\var$.

The semantics of a program ${\cal P}$ is determined in \crwl{} by
means of a proof calculus (see Fig.~\ref{fig:crwl}) for deriving
reduction statements $\prog \vdash e \clto t$, with $e \in Exp_\perp$
and $t \in CTerm_\perp$, meaning informally that $t$ is (or
approximates) a \emph{possible value} of $e$, obtained by iterated
reduction of $e$ using ${\cal P}$ under call-time choice.
\begin{figure}[tbp]
\begin{center}
  \framebox[.95\textwidth]{
\begin{minipage}{.9\textwidth}
\begin{center}
\begin{small}
\begin{tabular}{ll}
\regla{RR}{X \clto X}{}{$X \in \var$} &
\hspace{-.75cm}\regla{B}{e \clto \perp}{}{} \\
\regla{DC}{c(e_1, \ldots, e_n) \clto c(t_1, \ldots, t_n)}{e_1 \clto t_1 \ldots e_n \clto t_n}{$c \in {CS}^n$} \\
\regla{OR}{f(e_1, \ldots, e_n) \clto t}{e_1 \clto p_1\theta \ldots e_n \clto p_n\theta ~~r\theta \clto t}{$
      \begin{array}{l}
        f(p_1, \ldots, p_n)\tor r \in \prog\\
        \theta\in CSubst_\perp
      \end{array}$} \\
\end{tabular}
\end{small}
\end{center}
\end{minipage}
}
\end{center}
    \caption{Rules of \crwl}
    \label{fig:crwl}
\end{figure}
Rule \crule{B} (bottom) allows us to avoid the evaluation of any
expression, in order to get a non-strict semantics. Rules \crule{RR}
(restricted reflexivity) and \crule{DC} (decomposition) allow us to
reduce any variable to itself, and to decompose the evaluation of an
expression whose root symbol is a constructor. Rule \crule{OR} (outer
reduction) expresses that to evaluate a function call we must first
evaluate its arguments to get an instance of a program rule, perform
parameter passing (by means of a $CSubst_\perp$ $\theta$) and then
reduce the instantiated right-hand side. The use of partial
c-substitutions in {OR} is essential to express call-time choice, as
only single partial values are used for parameter passing. Notice also
that by the effect of $\theta$ in \crule{OR} extra variables in the
right-hand side of a rule can be replaced by any c-term, but not by
any expression.
%
The \emph{CRWL-denotation} of an expression $e \in Exp_\perp$ is
defined as $\den{e}^{\mathcal{P}}=\{t\in CTerm_\perp \mid \mathcal{P} \conscrwl e\crwlto t\}$.


\longText{\subsection{Isabelle/HOL}\label{isabelle}
\input{isabelle}}{}

\section{Formalizing \crwl\ in Isabelle}\label{formalizing}
\subsection{Basic definitions}
\label{sec:formalizing:basic}

We describe our formalization of \crwl{} in Isabelle. The first step
is to define elementary types for the syntactic elements.

\medskip

\begin{minipage}{\linewidth}
\begin{isacode}
\isacommand{datatype}\isamarkupfalse%
\ signat\ {\isacharequal}\ fs\ string\ %
{\isacharbar}\ cs\ string\ \ %
\isanewline
\isacommand{datatype}\isamarkupfalse%
\ varId\ {\isacharequal}\ vi\ string\isanewline
\isacommand{datatype}\isamarkupfalse%
\ exp\ {\isacharequal}\ perp\ {\isacharbar}\ Var\ varId\ 
{\isacharbar}\ Ap\ signat\ {\isachardoublequoteopen}exp\ list{\isachardoublequoteclose}
\isanewline
\isacommand{types}\isamarkupfalse%
\ \isanewline
\ \ subst\ {\isacharequal}\ {\isachardoublequoteopen}varId\ {\isasymRightarrow}\ exp\ option{\isachardoublequoteclose}\isanewline
\ \ rule\ {\isacharequal}\ {\isachardoublequoteopen}exp\ {\isacharasterisk}\ exp{\isachardoublequoteclose}\isanewline
\ \ program\ {\isacharequal}\ {\isachardoublequoteopen}rule\ set{\isachardoublequoteclose}
\end{isacode}
\end{minipage}

\medskip

\noindent%
Signatures are represented by a datatype that provides two
constructors \isaelem{cs} and \isaelem{fs} to distinguish between
constructor and function symbols.
%
%
The type \isaelem{varId} is used to represent variable identifiers,
which will be employed to define substitutions. Then the datatype
\isaelem{exp} is naturally defined following the inductive scheme of
$Exp_\perp$, therefore with this representation every expression is
partial by default.

Substitutions (type \isaelem{subst}) are represented as partial
functions from variable identifiers to expressions, using Isabelle's
\isaelem{option} type. Hence the domain of a substitution
{\isasymtheta} will be the set of elements from \isaelem{varId} for
which {\isasymtheta} returns some value different from \isaelem{None}.
Note that this representation does not ensure that domains of
substitutions are finite. 
Our proofs do not rely on this finiteness assumption.
%
Finally we represent a program rule as a pair of expressions, where
the first element is considered the left-hand side of the rule and the
second the right-hand side, and a program
simply as a set of program rules. The set of valid \crwl{} programs is
characterized by a predicate \isaelem{crwlProgram ::
  \isachardoublequoteopen program \ \isasymRightarrow \
  bool\isachardoublequoteclose}
that checks whether the restrictions of left-linearity and constructor
discipline are satisfied.

We define a function
\isaelem{apSubst ::
  \isachardoublequoteopen subst \isasymRightarrow\ exp
  \isasymRightarrow\ exp\isachardoublequoteclose}
for applying a substitution to an expression. The composition of
substitutions is defined through a function
\isaelem{substComp :: \isachardoublequoteopen
  subst \isasymRightarrow\ subst \isasymRightarrow\
  subst\isachardoublequoteclose}.
The following lemma ensures the correctness of this definition.

\medskip

\begin{minipage}{\linewidth}
\begin{isacode}
\isacommand{lemma}\isamarkupfalse%
\ subsCompAp\ {\isacharcolon}\\
\ \ {\isachardoublequoteopen}{\isacharparenleft}apSubst\ {\isasymtheta}\ {\isacharparenleft}apSubst\ {\isasymsigma}\ e{\isacharparenright}{\isacharparenright}\ {\isacharequal}\ {\isacharparenleft}apSubst\ {\isacharparenleft}substComp\ {\isasymtheta}\ {\isasymsigma}{\isacharparenright}\ e{\isacharparenright}{\isachardoublequoteclose}
\end{isacode}
\end{minipage}

\medskip

\noindent%
Just as ML, the Isabelle type system does not support subtyping, which
could have been useful to represent the sets of c-terms and
c-substitutions. Instead, we define predicates \isaelem{cterm} and
\isaelem{csubst} characterizing these
subtypes. 
We prove the expected lemmas, such as that the composition of two
c-substitutions is a c-substitution, or that the application of a
c-substitution to a c-term yields a c-term.

\subsection{Approximation order and contexts}\label{subsect:ordapContx}

Two key notions of \crwl{} have not yet been formalized: the
approximation order $\ordap$, which will be used in the formulation of
the polarity of \crwl, and the notion of one-hole context, which will
be used in the compositionality.

The following inductively defined predicate \isaelem{ordap} (with
concrete infix syntax \isaelem{\isasymsqsubseteq}) models the
approximation order.

\medskip

%
\begin{minipage}{\linewidth}
\begin{isacode}
\isacommand{inductive}\isamarkupfalse%
\isanewline
\ \ ordap\ {\isacharcolon}{\isacharcolon}\ {\isachardoublequoteopen}exp\ {\isasymRightarrow}\ exp\ {\isasymRightarrow}\ bool{\isachardoublequoteclose}\ {\isacharparenleft}{\isachardoublequoteopen}{\isacharunderscore}\ {\isasymsqsubseteq}\ {\isacharunderscore}{\isachardoublequoteclose}\ {\isacharbrackleft}{\isadigit{5}}{\isadigit{1}}{\isacharcomma}{\isadigit{5}}{\isadigit{1}}{\isacharbrackright}\ {\isadigit{5}}{\isadigit{0}}{\isacharparenright}\isanewline
\isakeyword{where}\isanewline
\ \ B
{\isacharcolon}\ {\isachardoublequoteopen}perp\ {\isasymsqsubseteq}\ e{\isachardoublequoteclose}\isanewline
{\isacharbar}\ V
{\isacharcolon}\ {\isachardoublequoteopen}Var\ x\ {\isasymsqsubseteq}\ Var\ x{\isachardoublequoteclose}\isanewline
{\isacharbar}\ Ap
{\isacharcolon}\ {\isachardoublequoteopen}{\isasymlbrakk}\ size\ es\ {\isacharequal}\ size\ es{\isacharprime}\ {\isacharsemicolon}\ ALL\ i\ {\isacharless}\ size\ es{\isachardot}\ es{\isacharbang}i\ {\isasymsqsubseteq}\ es{\isacharprime}{\isacharbang}i\ {\isasymrbrakk}\isanewline
\ \ \ \ \ \ \ {\isasymLongrightarrow}\ Ap\ h\ es\ {\isasymsqsubseteq}\ Ap\ h\ es{\isacharprime}{\isachardoublequoteclose}
\end{isacode}
\end{minipage}

\medskip

\noindent%
Rule \isaelem{B} asserts that \isaelem{perp \isasymsqsubseteq\ e}
holds for every \isaelem{e}; rule \isaelem{V} is needed for
\isaelem{\isasymsqsubseteq} to be reflexive; finally rule \isaelem{Ap}
ensures closedness under $\Sigma$-operations, and thus compatibility
with context~\cite{BaaderNipkow-98}, because
\isaelem{\isasymsqsubseteq} is reflexive and transitive, as we will
see. The following results state that our formulation of
\isaelem{\isasymsqsubseteq} defines a partial order.

\medskip

\begin{minipage}{\linewidth}
\begin{isacode}
\isacommand{lemma}\isamarkupfalse%
\ ordapRefl\ {\isacharcolon}\ {\isachardoublequoteopen}e\ {\isasymsqsubseteq}\ e{\isachardoublequoteclose}
 \isanewline
\isacommand{lemma}\isamarkupfalse%
\ ordapTrans\ {\isacharcolon}\isanewline
\ \ \isakeyword{assumes}\ {\isachardoublequoteopen}e{\isadigit{1}}\ {\isasymsqsubseteq}\ \ e{\isadigit{2}}{\isachardoublequoteclose}\ \isakeyword{and}\ {\isachardoublequoteopen}e{\isadigit{2}}\ {\isasymsqsubseteq}\ e{\isadigit{3}}{\isachardoublequoteclose}\isanewline
\ \ \isakeyword{shows}\ {\isachardoublequoteopen}e{\isadigit{1}}\ {\isasymsqsubseteq}\ \ e{\isadigit{3}}{\isachardoublequoteclose}
\isanewline
\isacommand{lemma}\isamarkupfalse%
\ ordapAntisym\ {\isacharcolon}\isanewline
\ \ \isakeyword{assumes}\ {\isachardoublequoteopen}e{\isadigit{1}}\ {\isasymsqsubseteq}\ e{\isadigit{2}}{\isachardoublequoteclose}\ \isakeyword{and}\ {\isachardoublequoteopen}e{\isadigit{2}}\ {\isasymsqsubseteq}\ e{\isadigit{1}}{\isachardoublequoteclose}\isanewline
\ \ \isakeyword{shows}\ {\isachardoublequoteopen}e{\isadigit{1}}\ {\isacharequal}\ e{\isadigit{2}}{\isachardoublequoteclose}\isanewline
\isacommand{definition}\isamarkupfalse%
\ ordap{\isacharunderscore}less\ {\isacharparenleft}{\isachardoublequoteopen}{\isacharunderscore}\ {\isasymsqsubset}\ {\isacharunderscore}{\isachardoublequoteclose}\ {\isacharbrackleft}{\isadigit{5}}{\isadigit{1}}{\isacharcomma}{\isadigit{5}}{\isadigit{1}}{\isacharbrackright}\ {\isadigit{5}}{\isadigit{0}}{\isacharparenright}\ \isakeyword{where}\isanewline
\ \ {\isachardoublequoteopen}e\ {\isasymsqsubset}\ e{\isacharprime}\ {\isasymequiv}\ e\ {\isasymsqsubseteq}\ e{\isacharprime}\ {\isasymand}\ e\ {\isasymnoteq}\ e{\isacharprime}{\isachardoublequoteclose}\isanewline
\isacommand{interpretation}\isamarkupfalse%
\ exp\ {\isacharcolon}\ order\ {\isacharbrackleft}ordap\ ordap{\isacharunderscore}less{\isacharbrackright}
%
\end{isacode}
\end{minipage}
%

\medskip

\noindent%
Contexts are represented as the datatype \isaelem{cntxt}, defined as follows:
%

\medskip

\begin{minipage}{\linewidth}
\begin{isacode}
\isacommand{datatype}\isamarkupfalse%
\ cntxt\ {\isacharequal}\ Hole\ {\isacharbar}\ Cperp\ {\isacharbar}\ CVar\ varId\ \isanewline
\ \ \ \ \ \ \ \ \ \ \ \ \ \ \ {\isacharbar}\ CAp\ signat\ {\isachardoublequoteopen}cntxt\ list{\isachardoublequoteclose}
\end{isacode}
\end{minipage}

\medskip

\noindent%
Note that \isaelem{cntxt} cannot follow the inductive structure of
$Cntxt$ with precision, because the type system of Isabelle is not
expressive enough to allow us to specify that only one of the
arguments of \isaelem{CAp} will be a context and the others will be
expressions.
Then our contexts are defined as expressions with possibly some holes inside.
Therefore the datatype \isaelem{cntxt} represents contexts with any
number of holes, even zero holes, and the function
\isaelem{apCon ::
  \isachardoublequoteopen exp \isasymRightarrow\ cntxt
  \isasymRightarrow\ exp\isachardoublequoteclose}
is defined so it puts the argument expression in every hole of the
argument context. In order to characterize contexts with just one
hole, we define a function
\isaelem{numHoles ::
  \isachardoublequoteopen cntxt \isasymRightarrow\
  nat\isachardoublequoteclose}
that returns the numbers of holes in a context. Using it we can define
define predicates \isaelem{oneHole} and \isaelem{noHole} and prove the
following lemmas.

\medskip

\begin{minipage}{\linewidth}
\begin{isacode}
\isacommand{lemma}\isamarkupfalse%
\ noHoleApDontCare\ {\isacharcolon}\isanewline
\ \ \isakeyword{assumes}\ {\isachardoublequoteopen}noHole\ xC{\isachardoublequoteclose}\isanewline
\ \ \isakeyword{shows}\ {\isachardoublequoteopen}apCon\ e\ xC\ {\isacharequal}\ apCon\ e{\isacharprime}\ xC{\isachardoublequoteclose}
\end{isacode}
\begin{isacode}
\isacommand{lemma}\isamarkupfalse%
\ oneHole\ {\isacharcolon}\isanewline
\ \ \isakeyword{assumes}\ {\isachardoublequoteopen}oneHole\ {\isacharparenleft}CAp\ h\ xCs{\isacharparenright}{\isachardoublequoteclose}\isanewline
\ \ \isakeyword{shows}\ {\isachardoublequoteopen}{\isasymexists}\ xC\ yCs\ zCs{\isachardot}\ xCs\ {\isacharequal}\ {\isacharparenleft}yCs\ {\isacharat}\ xC\ {\isacharhash}\ zCs{\isacharparenright}\ {\isasymand}\ oneHole\ xC\ {\isasymand}\isanewline
\ \ \ \ \ \ \ \ \ \ \ \ \ \ \ \ \ \ \ \ \ \ \ \
({\isasymforall}c\ {\isasymin}\ set\ {\isacharparenleft}yCs\ {\isacharat}\ zCs{\isacharparenright}.\ noHole\ c){\isachardoublequoteclose}
\end{isacode}
\end{minipage}

\subsection{The \crwl\ logic in Isabelle/HOL}
%
%
%

The \crwl\ logic has been formalized through the inductive predicate
\isaelem{clto} 
with infix notation
\isaelem{{\isachardoublequoteopen}{\isacharunderscore}\
  {\isasymturnstile}\ {\isacharunderscore}\ {\isasymrightarrow}\
  {\isacharunderscore}{\isachardoublequoteclose}}. The rules defining
\isaelem{clto} faithfully follow the inductive structure of the
definition of \crwl\ as it is presented in Fig.~\ref{fig:crwl}.

\medskip

%
\begin{minipage}{\linewidth} 
\begin{isacode}
\isacommand{inductive}\isamarkupfalse%
\ \isanewline
\ \ clto\ {\isacharcolon}{\isacharcolon}\ {\isachardoublequoteopen}program\ {\isasymRightarrow}\ exp\ {\isasymRightarrow}\ exp\ {\isasymRightarrow}\ bool{\isachardoublequoteclose}\ \ {\isacharparenleft}{\isachardoublequoteopen}{\isacharunderscore}\ {\isasymturnstile}\ {\isacharunderscore}\ {\isasymrightarrow}\ {\isacharunderscore}{\isachardoublequoteclose}\ {\isacharbrackleft}{\isadigit{1}}{\isadigit{0}}{\isadigit{0}}{\isacharcomma}{\isadigit{5}}{\isadigit{0}}{\isacharcomma}{\isadigit{5}}{\isadigit{0}}{\isacharbrackright}\ {\isadigit{3}}{\isadigit{8}}{\isacharparenright}\isanewline
\isakeyword{where}\isanewline
\ \ B{\isacharbrackleft}intro{\isacharbrackright}{\isacharcolon}\ \ {\isachardoublequoteopen}prog\ {\isasymturnstile}\ exp\ {\isasymrightarrow}\ perp{\isachardoublequoteclose}\isanewline
{\isacharbar}\ RR{\isacharbrackleft}intro{\isacharbrackright}{\isacharcolon}\ {\isachardoublequoteopen}prog\ {\isasymturnstile}\ Var\ v\ {\isasymrightarrow}\ Var\ v{\isachardoublequoteclose}\isanewline
{\isacharbar}\ DC{\isacharbrackleft}intro{\isacharbrackright}{\isacharcolon}\ {\isachardoublequoteopen}{\isasymlbrakk}size\ es\ {\isacharequal}\ size\ ts{\isacharsemicolon}\isanewline
\ \ \ \ \ \ \ \ \ \ \ \ \ \ \ {\isasymforall}i\ {\isacharless}\ size\ es{\isachardot}\ prog\ {\isasymturnstile}\ es{\isacharbang}i\ {\isasymrightarrow}\ ts{\isacharbang}i\isanewline
\ \ \ \ \ \ \ \ \ \ \ \ \ \ {\isasymrbrakk} {\isasymLongrightarrow}\ prog\ {\isasymturnstile}\ Ap\ {\isacharparenleft}cs\ c{\isacharparenright}\ es\ {\isasymrightarrow}\ Ap\ {\isacharparenleft}cs\ c{\isacharparenright}\ ts{\isachardoublequoteclose}\isanewline
{\isacharbar}\ OR{\isacharbrackleft}intro{\isacharbrackright}{\isacharcolon}\ {\isachardoublequoteopen}{\isasymlbrakk}{\isacharparenleft}Ap\ {\isacharparenleft}fs\ f{\isacharparenright}\ ps{\isacharcomma}\ r{\isacharparenright}\ {\isasymin}\ prog\ {\isacharsemicolon}\ csubst\ {\isasymtheta}\ {\isacharsemicolon}\isanewline
\ \ \ \ \ \ \ \ \ \ \ \ \ \ \ size\ es\ {\isacharequal}\ size\ ps\ {\isacharsemicolon}\isanewline
\ \ \ \ \ \ \ \ \ \ \ \ \ \ \ {\isasymforall}i\ {\isacharless}\ size\ es{\isachardot}\ prog\ {\isasymturnstile}\ es{\isacharbang}i\ {\isasymrightarrow}\ apSubst\ {\isasymtheta}\ {\isacharparenleft}ps{\isacharbang}i{\isacharparenright}{\isacharsemicolon}\isanewline
\ \ \ \ \ \ \ \ \ \ \ \ \ \ \ prog\ {\isasymturnstile}\ apSubst\ {\isasymtheta}\ r\ {\isasymrightarrow}\ t\isanewline
\ \ \ \ \ \ \ \ \ \ \ \ \ \ {\isasymrbrakk}\ {\isasymLongrightarrow}\ prog\ {\isasymturnstile}\ Ap\ {\isacharparenleft}fs\ f{\isacharparenright}\ es\ {\isasymrightarrow}\ t{\isachardoublequoteclose}
\end{isacode}
\end{minipage}

\medskip

\noindent%
Using \isaelem{clto} we can easily define the \crwl\ denotations in
Isabelle as follows.


\medskip

\begin{minipage}{\linewidth}
\begin{isacode}
\isacommand{definition}\isamarkupfalse%
\ den\ {\isacharcolon}{\isacharcolon}\ {\isachardoublequoteopen}program\ {\isasymRightarrow}\ exp\ {\isasymRightarrow}\ exp\ set{\isachardoublequoteclose}\ \isakeyword{where}\isanewline
\ \ {\isachardoublequoteopen}den\ P\ e\ {\isacharequal}\ {\isacharbraceleft}t{\isachardot}\ P\ {\isasymturnstile}\ e\ {\isasymrightarrow}\ t{\isacharbraceright}{\isachardoublequoteclose}
\end{isacode}
\end{minipage}


\section{Reasoning about \crwl\ in Isabelle}\label{properties}
%
\longText{
\indent Using the elements defined in the previous section, Isabelle is already able to prove some simple lemmas automatically, like the following reflexivity of \crwl\ for values.
\begin{isacode}
\isacommand{lemma}\isamarkupfalse%
\ ctermRefl\ {\isacharcolon}\ \isakeyword{assumes}\ {\isachardoublequoteopen}cterm\ t{\isachardoublequoteclose}\ \isakeyword{shows}\ {\isachardoublequoteopen}prog\ {\isasymturnstile}\ t\ {\isasymrightarrow}\ t{\isachardoublequoteclose}\isanewline
\isadelimproof
\endisadelimproof
\isatagproof
\isacommand{using}\isamarkupfalse%
\ assms\isanewline
\isacommand{by}\isamarkupfalse%
\ {\isacharparenleft}induct\ t\ rule{\isacharcolon}\ cterm{\isachardot}induct{\isacharcomma}\ auto{\isacharparenright}%
\endisatagproof
{\isafoldproof}%
%
%
\end{isacode}

\noindent Or the property that every value computed by \crwl\ is a partial c-term. 
\begin{isacode}
\isacommand{lemma}\isamarkupfalse%
\ ctermVals\ {\isacharcolon}\ \isakeyword{assumes}\ {\isachardoublequoteopen}P\ {\isasymturnstile}\ e\ {\isasymrightarrow}\ t{\isachardoublequoteclose}\ \isakeyword{shows}\ {\isachardoublequoteopen}cterm\ t{\isachardoublequoteclose}\isanewline
\isadelimproof
\endisadelimproof
\isatagproof
\isacommand{using}\isamarkupfalse%
\ assms\isanewline
\isacommand{by}\isamarkupfalse%
\ {\isacharparenleft}induct\ arbitrary{\isacharcolon}\ e{\isacharcomma}\ auto\ simp\ add{\isacharcolon}\ list{\isacharunderscore}all{\isacharunderscore}length\ list{\isacharunderscore}ball{\isacharunderscore}code{\isacharparenright}%
\endisatagproof
{\isafoldproof}%
%
\end{isacode}

\noindent And even more complex properties, like the following relating two inductive notions like \isaelem{clto} and \isaelem{ordap}. 
\begin{isacode}
\isacommand{lemma}\isamarkupfalse%
\ lessCTermOrdapL\ {\isacharcolon}\ \isakeyword{assumes}\ {\isachardoublequoteopen}P\ {\isasymturnstile}\ t\ {\isasymrightarrow}\ s{\isachardoublequoteclose}\ \isakeyword{and}\ {\isachardoublequoteopen}cterm\ t{\isachardoublequoteclose}\ \isakeyword{shows}\ {\isachardoublequoteopen}s\ {\isasymsqsubseteq}\ t{\isachardoublequoteclose}\isanewline
\isadelimproof
\endisadelimproof
\isatagproof
\isacommand{using}\isamarkupfalse%
\ assms\isanewline
\isacommand{by}\isamarkupfalse%
\ {\isacharparenleft}induct{\isacharcomma}\ auto{\isacharparenright}%
\endisatagproof
{\isafoldproof}%
%
\end{isacode}
%

}{}
The first interesting property that we are proving about \crwl{}
expresses that evaluation is \emph{closed under c-substitutions}:
reductions are preserved when terms are instantiated by
c-substitutions.

\medskip

\begin{minipage}{\linewidth}
\begin{isacode}
\isacommand{theorem}\isamarkupfalse%
\ crwlClosedCSubst\ {\isacharcolon}\ \isanewline
\ \ \isakeyword{assumes}\ {\isachardoublequoteopen}prog\ {\isasymturnstile}\ e\ {\isasymrightarrow}\ t{\isachardoublequoteclose}\ \isakeyword{and}\ {\isachardoublequoteopen}csubst\ {\isasymtheta}{\isachardoublequoteclose}\isanewline
\ \ \isakeyword{shows}\ {\isachardoublequoteopen}prog\ {\isasymturnstile}\ apSubst\ {\isasymtheta}\ e\ {\isasymrightarrow}\ apSubst\ {\isasymtheta}\ t{\isachardoublequoteclose}
\end{isacode}
\end{minipage}

\medskip

\noindent%
The proof of this lemma proceeds by induction on the \crwl-proof of the hypothesis, therefore we will have one case for each \crwl\ rule. The first three cases are proved automatically\longText{, although the case for \isaelem{RR} needs some tuning of the simplifier}{}. 
\longText{
\begin{isacode}
\isatagproof
\isacommand{using}\isamarkupfalse%
\ assms\isanewline
\isacommand{proof}\isamarkupfalse%
\ induct\isanewline
\ \ \isacommand{case}\isamarkupfalse%
\ B\ \isacommand{thus}\isamarkupfalse%
\ {\isacharquery}case\ \isacommand{by}\isamarkupfalse%
\ auto\isanewline
\isacommand{next}\isamarkupfalse%
\isanewline
\ \ \isacommand{case}\isamarkupfalse%
\ {\isacharparenleft}RR\ prog\ v{\isacharparenright}\ \isacommand{thus}\isamarkupfalse%
\ {\isacharquery}case\ \isacommand{by}\isamarkupfalse%
\ {\isacharparenleft}simp\ add{\isacharcolon}\ CSubsCterm\ ctermRefl\ del{\isacharcolon}\ apSubst{\isachardot}simps{\isacharparenright}\isanewline
\isacommand{next}\isamarkupfalse%
\isanewline
\ \ \isacommand{case}\isamarkupfalse%
\ {\isacharparenleft}DC\ es\ ts\ prog\ c{\isacharparenright}\ \isacommand{thus}\isamarkupfalse%
\ {\isacharquery}case\ \isacommand{by}\isamarkupfalse%
\ auto\isanewline
\isacommand{next}\isamarkupfalse%
\end{isacode}}{}
However, to prove the case for rule \isaelem{OR} Isabelle needs some
help from us. We need to prove
\[
\isaelem{
  prog\ {\isasymturnstile}\ {\isacharparenleft}Ap\ {\isacharparenleft}fs\ f{\isacharparenright}\ {\isacharparenleft}map\ {\isacharparenleft}apSubst\ {\isasymtheta}{\isacharparenright}\ es{\isacharparenright}{\isacharparenright}\ {\isasymrightarrow}\ {\isacharparenleft}apSubst\ {\isasymtheta}\ t{\isacharparenright}
}
\]
and then let the simplifier apply the definition of \isaelem{apSubst}.
In the proof for that subgoal we 
used lemma \isaelem{CSubsComp} to ensure that the c-substitution
\isaelem{{\isasymmu}} used for parameter passing composed with the
c-substitution \isaelem{{\isasymtheta}} in the hypothesis yields
another c-substitution, and lemma \isaelem{subsCompAp} to guarantee the
correct behaviour of the composition for those c-substitutions.
\longText{

\begin{isacode}
\ \ \isacommand{case}\isamarkupfalse%
\ {\isacharparenleft}OR\ f\ ps\ r\ prog\ {\isasymmu}\ es\ t{\isacharparenright}\isanewline
\ \ \ \ \isacommand{from}\isamarkupfalse%
\ OR\ \isacommand{have}\isamarkupfalse%
\ {\isachardoublequoteopen}prog\ {\isasymturnstile}\ {\isacharparenleft}Ap\ {\isacharparenleft}fs\ f{\isacharparenright}\ {\isacharparenleft}map\ {\isacharparenleft}apSubst\ {\isasymtheta}{\isacharparenright}\ es{\isacharparenright}{\isacharparenright}\ {\isasymrightarrow}\ {\isacharparenleft}apSubst\ {\isasymtheta}\ t{\isacharparenright}{\isachardoublequoteclose}\isanewline
\ \ \ \ \isacommand{proof}\isamarkupfalse%
\ {\isacharminus}\ %
\isanewline
\ \ \ \ \ \ \isacommand{from}\isamarkupfalse%
\ OR\ {\isacharparenleft}{\isadigit{2}}{\isacharparenright}\ \isakeyword{and}\ OR\ {\isacharparenleft}{\isadigit{7}}{\isacharparenright}\ \isakeyword{and}\ CSubsComp\ \isacommand{have}\isamarkupfalse%
\ {\isachardoublequoteopen}csubst\ {\isacharparenleft}substComp\ {\isasymtheta}\ {\isasymmu}{\isacharparenright}{\isachardoublequoteclose}\ \isacommand{by}\isamarkupfalse%
\ simp\isanewline
\ \ \ \ \ \isacommand{moreover}\isamarkupfalse%
\isanewline
\ \ \ \ \ \ \isacommand{from}\isamarkupfalse%
\ OR\ {\isacharparenleft}{\isadigit{3}}{\isacharparenright}\ \isacommand{have}\isamarkupfalse%
\ {\isachardoublequoteopen}size\ {\isacharparenleft}map\ {\isacharparenleft}apSubst\ {\isasymtheta}{\isacharparenright}\ es{\isacharparenright}\ {\isacharequal}\ size{\isacharparenleft}ps{\isacharparenright}{\isachardoublequoteclose}\ \isacommand{by}\isamarkupfalse%
\ simp\isanewline
\ \ \ \ \ \isacommand{moreover}\isamarkupfalse%
\isanewline
\ \ \ \ \ \ \isacommand{from}\isamarkupfalse%
\ OR\ \ \isacommand{have}\isamarkupfalse%
\ {\isachardoublequoteopen}ALL\ i{\isacharless}\ size\ {\isacharparenleft}map\ {\isacharparenleft}apSubst\ {\isasymtheta}{\isacharparenright}\ es{\isacharparenright}{\isachardot}\ clto\ prog\ {\isacharparenleft}{\isacharparenleft}map\ {\isacharparenleft}apSubst\ {\isasymtheta}{\isacharparenright}\ es{\isacharparenright}\ {\isacharbang}\ i{\isacharparenright}\ {\isacharparenleft}apSubst\ {\isacharparenleft}substComp\ {\isasymtheta}\ {\isasymmu}{\isacharparenright}\ {\isacharparenleft}ps\ {\isacharbang}\ i{\isacharparenright}{\isacharparenright}{\isachardoublequoteclose}\isanewline
\ \ \ \ \ \ \isacommand{proof}\isamarkupfalse%
\ {\isacharminus}\isanewline
\ \ \ \ \ \ \ \ \isacommand{from}\isamarkupfalse%
\ OR\ {\isacharparenleft}{\isadigit{4}}{\isacharparenright}\ \isakeyword{and}\ OR\ {\isacharparenleft}{\isadigit{7}}{\isacharparenright}\ \isacommand{have}\isamarkupfalse%
\ {\isachardoublequoteopen}ALL\ i\ {\isacharless}\ size\ es{\isachardot}\ clto\ prog\ {\isacharparenleft}apSubst\ {\isasymtheta}\ {\isacharparenleft}es\ {\isacharbang}\ i{\isacharparenright}{\isacharparenright}\ {\isacharparenleft}apSubst\ {\isasymtheta}\ {\isacharparenleft}apSubst\ {\isasymmu}\ {\isacharparenleft}ps\ {\isacharbang}\ i{\isacharparenright}{\isacharparenright}{\isacharparenright}{\isachardoublequoteclose}\ \isacommand{by}\isamarkupfalse%
\ simp\isanewline
\ \ \ \ \ \ \ \ \isacommand{with}\isamarkupfalse%
\ subsCompAp\ \isacommand{have}\isamarkupfalse%
\ {\isachardoublequoteopen}ALL\ i\ {\isacharless}\ size\ es{\isachardot}\ clto\ prog\ {\isacharparenleft}apSubst\ {\isasymtheta}\ {\isacharparenleft}es\ {\isacharbang}\ i{\isacharparenright}{\isacharparenright}\ {\isacharparenleft}apSubst\ {\isacharparenleft}substComp\ {\isasymtheta}\ {\isasymmu}{\isacharparenright}\ {\isacharparenleft}ps\ {\isacharbang}\ i{\isacharparenright}{\isacharparenright}{\isachardoublequoteclose}\ \isacommand{by}\isamarkupfalse%
\ simp\isanewline
\ \ \ \ \ \ \ \ \isacommand{thus}\isamarkupfalse%
\ {\isacharquery}thesis\ \isacommand{by}\isamarkupfalse%
\ simp\isanewline
\ \ \ \ \ \ \isacommand{qed}\isamarkupfalse%
\isanewline
\ \ \ \ \ \isacommand{moreover}\isamarkupfalse%
\isanewline
\ \ \ \ \ \ \isacommand{from}\isamarkupfalse%
\ OR\ \isacommand{have}\isamarkupfalse%
\ {\isachardoublequoteopen}clto\ prog\ {\isacharparenleft}apSubst\ {\isacharparenleft}substComp\ {\isasymtheta}\ {\isasymmu}{\isacharparenright}\ r{\isacharparenright}\ {\isacharparenleft}apSubst\ {\isasymtheta}\ t{\isacharparenright}{\isachardoublequoteclose}\isanewline
\ \ \ \ \ \ \isacommand{proof}\isamarkupfalse%
\ {\isacharminus}\isanewline
\ \ \ \ \ \ \ \ \isacommand{from}\isamarkupfalse%
\ OR\ {\isacharparenleft}{\isadigit{6}}{\isacharparenright}\ \isakeyword{and}\ OR\ {\isacharparenleft}{\isadigit{7}}{\isacharparenright}\ \isacommand{have}\isamarkupfalse%
\ {\isachardoublequoteopen}clto\ prog\ {\isacharparenleft}apSubst\ {\isasymtheta}\ {\isacharparenleft}apSubst\ {\isasymmu}\ r{\isacharparenright}{\isacharparenright}\ {\isacharparenleft}apSubst\ {\isasymtheta}\ t{\isacharparenright}{\isachardoublequoteclose}\ \isacommand{by}\isamarkupfalse%
\ simp\isanewline
\ \ \ \ \ \ \ \ \isacommand{with}\isamarkupfalse%
\ subsCompAp\ \isacommand{show}\isamarkupfalse%
\ {\isacharquery}thesis\ \isacommand{by}\isamarkupfalse%
\ simp\isanewline
\ \ \ \ \ \ \isacommand{qed}\isamarkupfalse%
\isanewline
\ \ \ \ \ \isacommand{ultimately}\isamarkupfalse%
\ \isacommand{show}\isamarkupfalse%
\ {\isacharquery}thesis\ \isacommand{using}\isamarkupfalse%
\ OR\ {\isacharparenleft}{\isadigit{1}}{\isacharparenright}\ \isacommand{by}\isamarkupfalse%
\ blast\isanewline
\ \ \ \ \isacommand{qed}\isamarkupfalse%
\isanewline
\ \ \ \ \isacommand{thus}\isamarkupfalse%
\ {\isacharquery}case\ \isacommand{by}\isamarkupfalse%
\ simp\isanewline
\isacommand{qed}\isamarkupfalse%
\endisatagproof
{\isafoldproof}%
\end{isacode}}{}

Note that for this result to hold no additional hypotheses about the
program or the expressions involved are needed. In particular, this
implies that the result holds even for programs that do not follow the
constructor discipline or that have non left-linear rules. The
Isabelle proof clearly shows that the important ingredients are the
use of c-substitutions for parameter passing and the reflexivity of
\crwl\ wrt. c-terms, expressed by lemma \isaelem{ctermRefl}, which
allows us to reduce to itself any expression \isaelem{X{\isasymtheta}}
coming from a premise \isaelem{X {\isasymrightarrow} X}.

\medskip

The second property that we address is the \emph{polarity of \crwl}.
This property is formulated by means of the approximation order and
roughly says that if we can compute a value for an expression then we
can compute a smaller value for a bigger expression. Here we should
understand the approximation order as an information order, in the
sense that the bigger the expression, the more information it gives,
and so more values can be computed from
it. 

\medskip

\begin{minipage}{\linewidth}
\begin{isacode}
\isacommand{theorem}\isamarkupfalse%
\ crwlPolarity\ {\isacharcolon}\ \isanewline
\ \ \isakeyword{assumes}\ {\isachardoublequoteopen}prog\ {\isasymturnstile}\ e\ {\isasymrightarrow}\ t{\isachardoublequoteclose}\ \isakeyword{and}\ {\isachardoublequoteopen}e\ {\isasymsqsubseteq}\ e{\isacharprime}{\isachardoublequoteclose}\ \isakeyword{and}\ {\isachardoublequoteopen}t{\isacharprime}\ {\isasymsqsubseteq}\ t{\isachardoublequoteclose}\isanewline
\ \ \isakeyword{shows}\ {\isachardoublequoteopen}prog\ {\isasymturnstile}\ e{\isacharprime}\ {\isasymrightarrow}\ t{\isacharprime}{\isachardoublequoteclose}
\isanewline
\isadelimproof
\endisadelimproof
\isatagproof
\isacommand{using}\isamarkupfalse%
\ assms\ 
\isacommand{proof}\isamarkupfalse%
\ {\isacharparenleft}induct\ arbitrary{\isacharcolon}\ e{\isacharprime}\ t{\isacharprime}{\isacharparenright}
\end{isacode}
\end{minipage}

\medskip

\noindent%
The idea of the proof is to construct a \crwl-proof for the conclusion
from the \crwl-proof of the hypothesis, hence it is natural to proceed
by induction on the structure of this proof (method \isaelem{induct}).
%
%
The qualifier \isaelem{arbitrary} is used to generalize the assertion
for any expressions \isaelem{e'} and \isaelem{t'}. The proof also
relies on the following additional lemmas about the approximation
order, which were proved automatically by Isabelle.

\medskip

\begin{minipage}{\linewidth}
\begin{isacode}
\isacommand{lemma}\isamarkupfalse%
\ ordapPerp{\isacharcolon}\ \isakeyword{assumes}\ {\isachardoublequoteopen}e\ {\isasymsqsubseteq}\ perp{\isachardoublequoteclose}\ \isakeyword{shows}\ {\isachardoublequoteopen}e\ {\isacharequal}\ perp{\isachardoublequoteclose}
\isanewline
\longText{
\isadelimproof
\endisadelimproof
\isatagproof
\isacommand{using}\isamarkupfalse%
\ assms\isanewline
\isacommand{by}\isamarkupfalse%
\ {\isacharparenleft}rule\ ordap{\isachardot}cases{\isacharcomma}\ auto{\isacharparenright}%
\endisatagproof
{\isafoldproof}%
\isadelimproof
\isanewline
\endisadelimproof
\isanewline}{}
\isacommand{lemma}\isamarkupfalse%
\ ordapVar{\isacharcolon}\ \isakeyword{assumes}\ {\isachardoublequoteopen}Var\ v\ {\isasymsqsubseteq}\ e{\isachardoublequoteclose}\ \isakeyword{shows}\ {\isachardoublequoteopen}e\ {\isacharequal}\ Var\ v{\isachardoublequoteclose}\isanewline
\longText{
\isadelimproof
\endisadelimproof
\isatagproof
\isacommand{using}\isamarkupfalse%
\ assms\ \isanewline
\isacommand{by}\isamarkupfalse%
\ {\isacharparenleft}rule\ ordap{\isachardot}cases{\isacharcomma}\ auto{\isacharparenright}%
\endisatagproof
{\isafoldproof}%
\isadelimproof
\isanewline
\endisadelimproof
\isanewline}{}
\isacommand{lemma}\isamarkupfalse%
\ ordapVar{\isacharunderscore}converse{\isacharcolon}\isanewline
\ \ \ \isakeyword{assumes}\ {\isachardoublequoteopen}e\ {\isasymsqsubseteq}\ Var\ v{\isachardoublequoteclose}\ \isakeyword{shows}\ {\isachardoublequoteopen}e\ {\isacharequal}\ perp\ {\isasymor}\ e\ {\isacharequal}\ Var\ v{\isachardoublequoteclose}\ \isanewline
\longText{
\isadelimproof
\endisadelimproof
\isatagproof
\isacommand{using}\isamarkupfalse%
\ assms\ \isanewline
\isacommand{by}\isamarkupfalse%
\ {\isacharparenleft}rule\ ordap{\isachardot}cases{\isacharcomma}\ auto{\isacharparenright}%
\endisatagproof
{\isafoldproof}%
\isadelimproof
\isanewline
\endisadelimproof
\isanewline}{}
\isacommand{lemma}\isamarkupfalse%
\ ordapAp{\isacharcolon}\isanewline
\ \ \ \isakeyword{assumes}\ {\isachardoublequoteopen}Ap\ h\ es\ {\isasymsqsubseteq}\ e{\isacharprime}{\isachardoublequoteclose}\isanewline
\ \ \ \isakeyword{shows}\ {\isachardoublequoteopen}{\isasymexists}es{\isacharprime}{\isachardot}\ e{\isacharprime}\ {\isacharequal}\ Ap\ h\ es{\isacharprime}\ {\isasymand}\ size\ es\ {\isacharequal}\ size\ es{\isacharprime}\isanewline
\ \ \ \ \ \ \ \ \ \ \ \ \ \ \ \ \ {\isasymand}\ {\isacharparenleft}ALL\ i\ {\isacharless}\ size\ es{\isachardot}\ es{\isacharbang}i\ {\isasymsqsubseteq}\ es{\isacharprime}{\isacharbang}i{\isacharparenright}{\isachardoublequoteclose}\isanewline
\longText{
\isadelimproof
\endisadelimproof
\isatagproof
\isacommand{using}\isamarkupfalse%
\ assms\isanewline
\isacommand{by}\isamarkupfalse%
\ {\isacharparenleft}rule\ ordap{\isachardot}cases{\isacharcomma}\ auto{\isacharparenright}%
\endisatagproof
{\isafoldproof}%
\isadelimproof
\isanewline
\endisadelimproof
\isanewline}{}
\isacommand{lemma}\isamarkupfalse%
\ ordapAp{\isacharunderscore}converse{\isacharcolon}\isanewline
\ \ \ \isakeyword{assumes}\ {\isachardoublequoteopen}e{\isacharprime}\ {\isasymsqsubseteq}\ Ap\ h\ es{\isachardoublequoteclose}\isanewline
\ \ \ \isakeyword{shows}\ {\isachardoublequoteopen}e{\isacharprime}\ {\isacharequal}\ perp\ {\isasymor}\isanewline
\ \ \ \ \ \ \ \ \ \ {\isacharparenleft}{\isasymexists}es{\isacharprime}{\isachardot}\ e{\isacharprime}\ {\isacharequal}\ Ap\ h\ es{\isacharprime}\ {\isasymand}\ size\ es\ {\isacharequal}\ size\ es{\isacharprime}\isanewline
\ \ \ \ \ \ \ \ \ \ \ \ \ \ \ \ \ \ {\isasymand}\ {\isacharparenleft}ALL\ i\ {\isacharless}\ size\ es{\isachardot}\ es{\isacharprime}{\isacharbang}i\ {\isasymsqsubseteq}\ es{\isacharbang}i{\isacharparenright}{\isacharparenright}{\isachardoublequoteclose}
\longText{
\isanewline
\isadelimproof
\endisadelimproof
\isatagproof
\isacommand{using}\isamarkupfalse%
\ assms\isanewline
\isacommand{by}\isamarkupfalse%
\ {\isacharparenleft}rule\ ordap{\isachardot}cases{\isacharcomma}\ auto{\isacharparenright}%
\endisatagproof
{\isafoldproof}}{}%
%
\end{isacode}
\end{minipage}

\medskip

\noindent%
The inductive proof for theorem~\isaelem{crwlPolarity} again considers
each \crwl{} rule in turn. In the case for \isaelem{B} we have
\isaelem{t = perp}, hence we just have to apply \isaelem{ordapPerp} to
get \isaelem{t' = perp}, and then use the \crwl\ rule \isaelem{B}.
\longText{
\begin{isacode}
\isacommand{case}\isamarkupfalse%
\ B\ \isacommand{thus}\isamarkupfalse%
\ {\isacharquery}case\isanewline
\ \ \ \ \isacommand{proof}\isamarkupfalse%
\ {\isacharminus}\isanewline
\ \ \ \ \ \ \isacommand{from}\isamarkupfalse%
\ B\ \isakeyword{and}\ ordapPerp\ \isacommand{have}\isamarkupfalse%
\ {\isachardoublequoteopen}t{\isacharprime}\ {\isacharequal}\ perp{\isachardoublequoteclose}\ \isacommand{by}\isamarkupfalse%
\ simp\isanewline
\ \ \ \ \ \ \isacommand{thus}\isamarkupfalse%
\ {\isacharquery}thesis\ \isacommand{by}\isamarkupfalse%
\ auto\isanewline
\ \ \ \ \isacommand{qed}\isamarkupfalse%
\isanewline
\isacommand{next}\isamarkupfalse%
\end{isacode}}{}
Regarding \isaelem{RR}, as then \isaelem{t = Var~v}, by
\isaelem{ordapVar{\isacharunderscore}converse} we get that either
\isaelem{t' = perp} or \isaelem{t' = Var~v}. The first case is
trivial, and in the latter we just have to apply \isaelem{ordapVar}
getting \isaelem{e' = Var~v}, which is enough for Isabelle to finish
the proof automatically.
\longText{

\begin{isacode}
\ \ \isacommand{case}\isamarkupfalse%
\ {\isacharparenleft}RR\ prog\ v\ e{\isacharprime}\ t{\isacharprime}{\isacharparenright}\ \isacommand{thus}\isamarkupfalse%
\ {\isacharquery}case\isanewline
\ \ \ \ \isacommand{proof}\isamarkupfalse%
\ {\isacharminus}\isanewline
\ \ \ \ \ \ \isacommand{from}\isamarkupfalse%
\ RR\ \isakeyword{and}\ ordapVar{\isacharunderscore}converse\ \isacommand{have}\isamarkupfalse%
\ {\isachardoublequoteopen}t{\isacharprime}\ {\isacharequal}\ perp\ {\isasymor}\ t{\isacharprime}\ {\isacharequal}\ {\isacharparenleft}Var\ v{\isacharparenright}{\isachardoublequoteclose}\ \isacommand{by}\isamarkupfalse%
\ simp\isanewline
\ \ \ \ \ \isacommand{moreover}\isamarkupfalse%
\isanewline
\ \ \ \ \ \ \isacommand{{\isacharbraceleft}}\isamarkupfalse%
\isanewline
\ \ \ \ \ \ \ \isacommand{assume}\isamarkupfalse%
\ {\isachardoublequoteopen}t{\isacharprime}\ {\isacharequal}\ perp{\isachardoublequoteclose}\ \isacommand{hence}\isamarkupfalse%
\ {\isacharquery}thesis\ \isacommand{by}\isamarkupfalse%
\ auto\ \isanewline
\ \ \ \ \ \ \isacommand{{\isacharbraceright}}\isamarkupfalse%
\isanewline
\ \ \ \ \ \isacommand{moreover}\isamarkupfalse%
\isanewline
\ \ \ \ \ \ \isacommand{{\isacharbraceleft}}\isamarkupfalse%
\isanewline
\ \ \ \ \ \ \ \isacommand{assume}\isamarkupfalse%
\ t{\isacharprime}Var\ {\isacharcolon}\ {\isachardoublequoteopen}t{\isacharprime}\ {\isacharequal}\ {\isacharparenleft}Var\ v{\isacharparenright}{\isachardoublequoteclose}\isanewline
\ \ \ \ \ \ \ \isacommand{from}\isamarkupfalse%
\ RR\ \isakeyword{and}\ ordapVar\ \isacommand{have}\isamarkupfalse%
\ {\isachardoublequoteopen}e{\isacharprime}\ {\isacharequal}\ {\isacharparenleft}Var\ v{\isacharparenright}{\isachardoublequoteclose}\ \isacommand{by}\isamarkupfalse%
\ simp\isanewline
\ \ \ \ \ \ \ \isacommand{with}\isamarkupfalse%
\ t{\isacharprime}Var\ \isacommand{have}\isamarkupfalse%
\ {\isacharquery}thesis\ \isacommand{by}\isamarkupfalse%
\ auto\isanewline
\ \ \ \ \ \ \isacommand{{\isacharbraceright}}\isamarkupfalse%
\isanewline
\ \ \ \ \ \isacommand{ultimately}\isamarkupfalse%
\ \isacommand{show}\isamarkupfalse%
\ {\isacharquery}thesis\ \isacommand{by}\isamarkupfalse%
\ auto\isanewline
\ \ \ \ \isacommand{qed}\isamarkupfalse%
\isanewline
\isacommand{next}\isamarkupfalse%
\end{isacode}}{}
The case of \isaelem{DC} is more complicated. Again we obtain two
cases for \isaelem{t' = perp} and \isaelem{t'} a constructor
application, by using lemma
\isaelem{ordapAp{\isacharunderscore}converse}. While the first case is
trivial, the second one requires some involved reasoning over the list
of arguments, using the information we get from applying lemma
\isaelem{ordapAp}.
%
\longText{

\begin{isacode}
\ \ \isacommand{case}\isamarkupfalse%
\ {\isacharparenleft}DC\ es\ ts\ prog\ c\ e{\isacharprime}\ t{\isacharprime}{\isacharparenright}\ \isacommand{thus}\isamarkupfalse%
\ {\isacharquery}case\isanewline
\ \ \ \ \isacommand{proof}\isamarkupfalse%
\ {\isacharminus}\isanewline
\ \ \ \ \ \ \isacommand{from}\isamarkupfalse%
\ DC\ \isakeyword{and}\ ordapAp{\isacharunderscore}converse\ \isacommand{have}\isamarkupfalse%
\ {\isachardoublequoteopen}{\isacharparenleft}t{\isacharprime}\ {\isacharequal}\ perp{\isacharparenright}\ {\isasymor}\ {\isacharparenleft}{\isasymexists}\ ts{\isacharprime}\ {\isachardot}\ {\isacharparenleft}t{\isacharprime}\ {\isacharequal}\ Ap\ {\isacharparenleft}cs\ c{\isacharparenright}\ ts{\isacharprime}{\isacharparenright}\ {\isasymand}\ {\isacharparenleft}size\ ts\ {\isacharequal}\ size\ ts{\isacharprime}{\isacharparenright}\ {\isasymand}\ {\isacharparenleft}ALL\ i\ {\isacharless}\ {\isacharparenleft}size\ ts{\isacharparenright}{\isachardot}\ {\isacharparenleft}ts{\isacharprime}\ {\isacharbang}\ i{\isacharparenright}\ {\isasymsqsubseteq}\ {\isacharparenleft}ts\ {\isacharbang}\ i{\isacharparenright}{\isacharparenright}{\isacharparenright}{\isachardoublequoteclose}\ \isacommand{by}\isamarkupfalse%
\ simp\isanewline
\ \ \ \ \ \isacommand{moreover}\isamarkupfalse%
\isanewline
\ \ \ \ \ \ \isacommand{{\isacharbraceleft}}\isamarkupfalse%
\isanewline
\ \ \ \ \ \ \ \isacommand{assume}\isamarkupfalse%
\ {\isachardoublequoteopen}t{\isacharprime}\ {\isacharequal}\ perp{\isachardoublequoteclose}\ \isacommand{hence}\isamarkupfalse%
\ {\isacharquery}thesis\ \isacommand{by}\isamarkupfalse%
\ auto\isanewline
\ \ \ \ \ \ \isacommand{{\isacharbraceright}}\isamarkupfalse%
\ \isanewline
\ \ \ \ \ \isacommand{moreover}\isamarkupfalse%
\isanewline
\ \ \ \ \ \ \isacommand{{\isacharbraceleft}}\isamarkupfalse%
\isanewline
\ \ \ \ \ \ \ \isacommand{assume}\isamarkupfalse%
\ t{\isacharprime}ApC\ {\isacharcolon}{\isachardoublequoteopen}{\isasymexists}\ ts{\isacharprime}\ {\isachardot}\ {\isacharparenleft}t{\isacharprime}\ {\isacharequal}\ Ap\ {\isacharparenleft}cs\ c{\isacharparenright}\ ts{\isacharprime}{\isacharparenright}\ {\isasymand}\ {\isacharparenleft}size\ ts\ {\isacharequal}\ size\ ts{\isacharprime}{\isacharparenright}\ {\isasymand}\ {\isacharparenleft}ALL\ i\ {\isacharless}\ {\isacharparenleft}size\ ts{\isacharparenright}{\isachardot}\ {\isacharparenleft}ts{\isacharprime}\ {\isacharbang}\ i{\isacharparenright}\ {\isasymsqsubseteq}\ {\isacharparenleft}ts\ {\isacharbang}\ i{\isacharparenright}{\isacharparenright}{\isachardoublequoteclose}\isanewline
\ \ \ \ \ \ \ \isacommand{then}\isamarkupfalse%
\ \isacommand{obtain}\isamarkupfalse%
\ ts{\isacharprime}\ \isakeyword{where}\ t{\isacharprime}Shape\ {\isacharcolon}\ {\isachardoublequoteopen}{\isacharparenleft}t{\isacharprime}\ {\isacharequal}\ Ap\ {\isacharparenleft}cs\ c{\isacharparenright}\ ts{\isacharprime}{\isacharparenright}\ {\isasymand}\ {\isacharparenleft}size\ ts\ {\isacharequal}\ size\ ts{\isacharprime}{\isacharparenright}\ {\isasymand}\ {\isacharparenleft}ALL\ i\ {\isacharless}\ {\isacharparenleft}size\ ts{\isacharparenright}{\isachardot}\ {\isacharparenleft}ts{\isacharprime}\ {\isacharbang}\ i{\isacharparenright}\ {\isasymsqsubseteq}\ {\isacharparenleft}ts\ {\isacharbang}\ i{\isacharparenright}{\isacharparenright}{\isachardoublequoteclose}\ \isacommand{{\isachardot}{\isachardot}}\isamarkupfalse%
\isanewline
\ \ \ \ \ \ \ \isacommand{from}\isamarkupfalse%
\ DC\ {\isacharparenleft}{\isadigit{3}}{\isacharparenright}\ \isakeyword{and}\ ordapAp\ \isacommand{have}\isamarkupfalse%
\ {\isachardoublequoteopen}{\isasymexists}\ es{\isacharprime}\ {\isachardot}\ {\isacharparenleft}e{\isacharprime}\ {\isacharequal}\ Ap\ {\isacharparenleft}cs\ c{\isacharparenright}\ es{\isacharprime}{\isacharparenright}\ {\isasymand}\ {\isacharparenleft}size\ es\ {\isacharequal}\ size\ es{\isacharprime}{\isacharparenright}\ {\isasymand}\ {\isacharparenleft}ALL\ i\ {\isacharless}\ {\isacharparenleft}size\ es{\isacharparenright}{\isachardot}\ {\isacharparenleft}es\ {\isacharbang}\ i{\isacharparenright}\ {\isasymsqsubseteq}\ {\isacharparenleft}es{\isacharprime}\ {\isacharbang}\ i{\isacharparenright}{\isacharparenright}{\isachardoublequoteclose}\ \isacommand{by}\isamarkupfalse%
\ simp\isanewline
\ \ \ \ \ \ \ \isacommand{then}\isamarkupfalse%
\ \isacommand{obtain}\isamarkupfalse%
\ es{\isacharprime}\ \isakeyword{where}\ e{\isacharprime}Shape\ {\isacharcolon}\ {\isachardoublequoteopen}{\isacharparenleft}e{\isacharprime}\ {\isacharequal}\ Ap\ {\isacharparenleft}cs\ c{\isacharparenright}\ es{\isacharprime}{\isacharparenright}\ {\isasymand}\ {\isacharparenleft}size\ es\ {\isacharequal}\ size\ es{\isacharprime}{\isacharparenright}\ {\isasymand}\ {\isacharparenleft}ALL\ i\ {\isacharless}\ {\isacharparenleft}size\ es{\isacharparenright}{\isachardot}\ {\isacharparenleft}es\ {\isacharbang}\ i{\isacharparenright}\ {\isasymsqsubseteq}\ {\isacharparenleft}es{\isacharprime}\ {\isacharbang}\ i{\isacharparenright}{\isacharparenright}{\isachardoublequoteclose}\ \isacommand{{\isachardot}{\isachardot}}\isamarkupfalse%
\isanewline
\ \ \ \ \ \ \ \isacommand{with}\isamarkupfalse%
\ t{\isacharprime}Shape\ \isakeyword{and}\ DC\ {\isacharparenleft}{\isadigit{1}}{\isacharparenright}\ \isacommand{have}\isamarkupfalse%
\ size{\isacharprime}s\ {\isacharcolon}\ {\isachardoublequoteopen}size\ es{\isacharprime}\ {\isacharequal}\ size\ ts{\isacharprime}{\isachardoublequoteclose}\ \isacommand{by}\isamarkupfalse%
\ simp\isanewline
\ \ \ \ \ \ \ \isacommand{from}\isamarkupfalse%
\ size{\isacharprime}s\ \isakeyword{and}\ e{\isacharprime}Shape\ \isakeyword{and}\ t{\isacharprime}Shape\ \isakeyword{and}\ DC\ {\isacharparenleft}{\isadigit{2}}{\isacharparenright}\ \isacommand{have}\isamarkupfalse%
\ {\isachardoublequoteopen}ALL\ i\ {\isacharless}\ {\isacharparenleft}size\ es{\isacharparenright}{\isachardot}\ prog\ {\isasymturnstile}\ {\isacharparenleft}es{\isacharprime}\ {\isacharbang}\ i{\isacharparenright}\ {\isasymrightarrow}\ {\isacharparenleft}ts{\isacharprime}\ {\isacharbang}\ i{\isacharparenright}{\isachardoublequoteclose}\ \isacommand{by}\isamarkupfalse%
\ auto\isanewline
\ \ \ \ \ \ \ \isacommand{with}\isamarkupfalse%
\ size{\isacharprime}s\ \isakeyword{and}\ e{\isacharprime}Shape\ \isakeyword{and}\ t{\isacharprime}Shape\ \isacommand{have}\isamarkupfalse%
\ {\isacharquery}thesis\ \isacommand{by}\isamarkupfalse%
\ auto\isanewline
\ \ \ \ \ \ \isacommand{{\isacharbraceright}}\isamarkupfalse%
\isanewline
\ \ \ \ \ \isacommand{ultimately}\isamarkupfalse%
\ \isacommand{show}\isamarkupfalse%
\ {\isacharquery}thesis\ \isacommand{by}\isamarkupfalse%
\ auto\isanewline
\ \ \ \ \isacommand{qed}\isamarkupfalse%
\isanewline
\isacommand{next}\isamarkupfalse%
\end{isacode}}{}
Finally, the proof for \isaelem{OR} is similar to the second case of
the proof for \isaelem{DC}, with a similar manipulation of the list of
arguments, and the use of lemma \isaelem{ordapAp} to obtain the
induction hypothesis for the arguments.
\longText{

\begin{isacode}
\ \ \isacommand{case}\isamarkupfalse%
\ {\isacharparenleft}OR\ f\ ps\ r\ prog\ {\isasymtheta}\ es\ t\ e{\isacharprime}\ t{\isacharprime}{\isacharparenright}\ \isacommand{thus}\isamarkupfalse%
\ {\isacharquery}case\isanewline
\ \ \ \ \isacommand{proof}\isamarkupfalse%
\ {\isacharminus}\isanewline
\ \ \ \ \ \ \isacommand{from}\isamarkupfalse%
\ OR\ {\isacharparenleft}{\isadigit{7}}{\isacharparenright}\ \isakeyword{and}\ ordapAp\ \isacommand{have}\isamarkupfalse%
\ {\isachardoublequoteopen}{\isasymexists}\ es{\isacharprime}\ {\isachardot}\ {\isacharparenleft}e{\isacharprime}\ {\isacharequal}\ Ap\ {\isacharparenleft}fs\ f{\isacharparenright}\ es{\isacharprime}{\isacharparenright}\ {\isasymand}\ {\isacharparenleft}size\ es\ {\isacharequal}\ size\ es{\isacharprime}{\isacharparenright}\ {\isasymand}\ {\isacharparenleft}ALL\ i\ {\isacharless}\ {\isacharparenleft}size\ es{\isacharparenright}{\isachardot}\ {\isacharparenleft}es\ {\isacharbang}\ i{\isacharparenright}\ {\isasymsqsubseteq}\ {\isacharparenleft}es{\isacharprime}\ {\isacharbang}\ i{\isacharparenright}{\isacharparenright}{\isachardoublequoteclose}\ \isacommand{by}\isamarkupfalse%
\ simp\isanewline
\ \ \ \ \ \ \isacommand{then}\isamarkupfalse%
\ \isacommand{obtain}\isamarkupfalse%
\ es{\isacharprime}\ \isakeyword{where}\ e{\isacharprime}Shape\ {\isacharcolon}\ {\isachardoublequoteopen}{\isacharparenleft}e{\isacharprime}\ {\isacharequal}\ Ap\ {\isacharparenleft}fs\ f{\isacharparenright}\ es{\isacharprime}{\isacharparenright}\ {\isasymand}\ {\isacharparenleft}size\ es\ {\isacharequal}\ size\ es{\isacharprime}{\isacharparenright}\ {\isasymand}\ {\isacharparenleft}ALL\ i\ {\isacharless}\ {\isacharparenleft}size\ es{\isacharparenright}{\isachardot}\ {\isacharparenleft}es\ {\isacharbang}\ i{\isacharparenright}\ {\isasymsqsubseteq}\ {\isacharparenleft}es{\isacharprime}\ {\isacharbang}\ i{\isacharparenright}{\isacharparenright}{\isachardoublequoteclose}\ \isacommand{{\isachardot}{\isachardot}}\isamarkupfalse%
\isanewline
\ \ \ \ \ \ \isacommand{with}\isamarkupfalse%
\ OR\ {\isacharparenleft}{\isadigit{4}}{\isacharparenright}\ \isakeyword{and}\ ordapRefl\ \isacommand{have}\isamarkupfalse%
\ prems{\isacharprime}\ {\isacharcolon}\ {\isachardoublequoteopen}ALL\ i\ {\isacharless}\ {\isacharparenleft}size\ es{\isacharparenright}{\isachardot}\ prog\ {\isasymturnstile}\ {\isacharparenleft}es{\isacharprime}\ {\isacharbang}\ i{\isacharparenright}\ {\isasymrightarrow}\ apSubst\ {\isasymtheta}\ {\isacharparenleft}ps\ {\isacharbang}\ i{\isacharparenright}{\isachardoublequoteclose}\ \isacommand{by}\isamarkupfalse%
\ simp\isanewline
\ \ \ \ \ \ \isacommand{from}\isamarkupfalse%
\ OR\ {\isacharparenleft}{\isadigit{3}}{\isacharparenright}\ \isakeyword{and}\ e{\isacharprime}Shape\ \isacommand{have}\isamarkupfalse%
\ sizes{\isacharprime}\ {\isacharcolon}\ {\isachardoublequoteopen}size\ es{\isacharprime}\ {\isacharequal}\ size\ ps{\isachardoublequoteclose}\ \isacommand{by}\isamarkupfalse%
\ simp\ \isanewline
\ \ \ \ \ \ \isacommand{from}\isamarkupfalse%
\ OR\ {\isacharparenleft}{\isadigit{6}}{\isacharparenright}\ OR\ {\isacharparenleft}{\isadigit{8}}{\isacharparenright}\ \isakeyword{and}\ ordapRefl\ \isacommand{have}\isamarkupfalse%
\ {\isachardoublequoteopen}prog\ {\isasymturnstile}\ apSubst\ {\isasymtheta}\ r\ {\isasymrightarrow}\ t{\isacharprime}{\isachardoublequoteclose}\ \isacommand{by}\isamarkupfalse%
\ simp\isanewline
\ \ \ \ \ \ \isacommand{with}\isamarkupfalse%
\ OR\ {\isacharparenleft}{\isadigit{1}}{\isacharparenright}\ \isakeyword{and}\ OR\ {\isacharparenleft}{\isadigit{2}}{\isacharparenright}\ \isakeyword{and}\ sizes{\isacharprime}\ \isakeyword{and}\ prems{\isacharprime}\ \isakeyword{and}\ e{\isacharprime}Shape\ \isacommand{show}\isamarkupfalse%
\ {\isacharquery}thesis\ \isacommand{by}\isamarkupfalse%
\ auto\isanewline
\ \ \ \ \isacommand{qed}\isamarkupfalse%
\isanewline
\isacommand{qed}\isamarkupfalse%
\endisatagproof
{\isafoldproof}%
\isadelimproof
\isanewline
\endisadelimproof
\end{isacode}}{}

Once again we find that this proof does not require any hypothesis on
the linearity or the constructor discipline of the program: this is
indeed quite obvious because this property only talks about what
happens when we replace some subexpression by \isaelem{perp}.

\medskip

Finally we will tackle the \emph{compositionality of \crwl}, that says that if we take a context with just one hole and an expression, then the set of values for the expression put it that context will be the union of the set of values for the result of putting each value for the expression in that context. 

\medskip

\begin{minipage}{\linewidth}
\begin{isacode}
\isacommand{theorem}\isamarkupfalse%
\ compCRWL\ {\isacharcolon}\ \isanewline
\ \ \ \isakeyword{assumes}\ {\isachardoublequoteopen}oneHole\ xC{\isachardoublequoteclose}\ \isanewline
\ \ \ \isakeyword{shows}\ {\isachardoublequoteopen}den\ P\ {\isacharparenleft}apCon\ e\ xC{\isacharparenright}\ {\isacharequal}\isanewline
\ \ \ \ \ \ \ \ \ \ \ {\isacharparenleft}{\isasymUnion}t{\isasymin}den\ P\ e{\isachardot}\ den\ P\ {\isacharparenleft}apCon\ t\ xC{\isacharparenright}{\isacharparenright}{\isachardoublequoteclose}
\end{isacode}
\end{minipage}

\medskip

\noindent%
We have proved the two set inclusions separately as auxiliary
lemmas\longText{.

\begin{isacode}
\isacommand{lemma}\isamarkupfalse%
\ compCRWL{\isadigit{1}}\ {\isacharcolon}\ \isanewline
\ \ \isakeyword{assumes}\ {\isachardoublequoteopen}P\ {\isasymturnstile}\ a\ {\isasymrightarrow}\ t{\isachardoublequoteclose}\
{\isachardoublequoteopen}a\ {\isacharequal}\ apCon\ e\ xC{\isachardoublequoteclose}\ \isakeyword{and}\ {\isachardoublequoteopen}oneHole\ xC{\isachardoublequoteclose}\isanewline
\ \ \isakeyword{shows}\ {\isachardoublequoteopen}{\isasymexists}\ s{\isachardot}\ P\ {\isasymturnstile}\ e\ {\isasymrightarrow}\ s\ {\isasymand}\ P\ {\isasymturnstile}\ apCon\ s\ xC\ {\isasymrightarrow}\ t{\isachardoublequoteclose}\isanewline
\isadelimproof
\endisadelimproof
\isatagproof
\isacommand{using}\isamarkupfalse%
\ assms\ \isacommand{unfolding}\isamarkupfalse%
\ oneHole{\isacharunderscore}def\isanewline
\isacommand{proof}\isamarkupfalse%
\ {\isacharparenleft}induct\ arbitrary{\isacharcolon}\ e\ xC{\isacharparenright}\isanewline

\isacommand{lemma}\isamarkupfalse%
\ compCRWL{\isadigit{2}}\ {\isacharcolon}\ \isanewline
\ \ \isakeyword{assumes}\ {\isachardoublequoteopen}P\ {\isasymturnstile}\ a\ {\isasymrightarrow}\ t{\isachardoublequoteclose}\
{\isachardoublequoteopen}a\ {\isacharequal}\ apCon\ s\ xC{\isachardoublequoteclose}\
{\isachardoublequoteopen}oneHole\ xC{\isachardoublequoteclose}\ \isakeyword{and}\ {\isachardoublequoteopen}P\ {\isasymturnstile}\ e\ {\isasymrightarrow}\ s{\isachardoublequoteclose}\isanewline
\ \ \isakeyword{shows}\ {\isachardoublequoteopen}P\ {\isasymturnstile}\ apCon\ e\ xC\ {\isasymrightarrow}\ t{\isachardoublequoteclose}\isanewline
\isadelimproof
\endisadelimproof
\isatagproof
\isacommand{using}\isamarkupfalse%
\ assms\ \isacommand{unfolding}\isamarkupfalse%
\ oneHole{\isacharunderscore}def\isanewline
\isacommand{proof}\isamarkupfalse%
\ {\isacharparenleft}induct\ arbitrary{\isacharcolon}\ s\ xC\ e{\isacharparenright}
\end{isacode}

}{ \isaelem{compCRWL1} and \isaelem{compCRWL2}.}
The proofs of these lemmas are quite laborious but essentially
proceed by induction on the \crwl-proof in their hypothesis, using it
to build a \crwl-proof for the statement in the conclusion. In these
proofs, Lemma \isaelem{noHoleApDontCare} from
Subsect.~\ref{subsect:ordapContx} is fundamental%
\longText{, as are the
  following versions of lemmas \isaelem{compCRWL1} and
  \isaelem{compCRWL2}, for the particular case of having a hole as
  context.
\begin{isacode}
\isacommand{lemma}\isamarkupfalse%
\ compCRWLHole{\isadigit{1}}\ {\isacharcolon}\ \isanewline
\ \ \isakeyword{assumes}\ {\isachardoublequoteopen}P\ {\isasymturnstile}\ apCon\ e\ xC\ {\isasymrightarrow}\ t{\isachardoublequoteclose}\ \isakeyword{and}\ {\isachardoublequoteopen}xC\ {\isacharequal}\ Hole{\isachardoublequoteclose}\isanewline
\ \ \isakeyword{shows}\ {\isachardoublequoteopen}{\isasymexists}\ t{\isacharprime}{\isachardot}\ P\ \ {\isasymturnstile}\ e\ {\isasymrightarrow}\ t{\isacharprime}\ {\isasymand}\ P\ {\isasymturnstile}\ apCon\ t{\isacharprime}\ xC\ {\isasymrightarrow}\ t{\isachardoublequoteclose}
\isadelimproof
\isanewline
\endisadelimproof
\isatagproof
\isacommand{proof}\isamarkupfalse%
\ {\isacharminus}\isanewline
\ \ \isacommand{from}\isamarkupfalse%
\ assms\ \isacommand{have}\isamarkupfalse%
\ derHole\ {\isacharcolon}\ {\isachardoublequoteopen}P\ {\isasymturnstile}\ e\ {\isasymrightarrow}\ t{\isachardoublequoteclose}\ \isacommand{by}\isamarkupfalse%
\ auto\isanewline
\ \ \isacommand{with}\isamarkupfalse%
\ ctermRefl\ \isakeyword{and}\ ctermVals\ \isacommand{have}\isamarkupfalse%
\ {\isachardoublequoteopen}P\ {\isasymturnstile}\ t\ {\isasymrightarrow}\ t{\isachardoublequoteclose}\ \isacommand{by}\isamarkupfalse%
\ simp\isanewline
\ \ \isacommand{with}\isamarkupfalse%
\ derHole\ \isakeyword{and}\ assms\ \isacommand{show}\isamarkupfalse%
\ {\isacharquery}thesis\ \isacommand{by}\isamarkupfalse%
\ auto\isanewline
\isacommand{qed}\isamarkupfalse%
\endisatagproof
{\isafoldproof}%
\end{isacode}

\begin{isacode}
\isacommand{lemma}\isamarkupfalse%
\ compCRWLHole{\isadigit{2}}\ {\isacharcolon}\ \isanewline
\ \ \isakeyword{assumes}\ {\isachardoublequoteopen}P\ {\isasymturnstile}\ apCon\ s\ Hole\ {\isasymrightarrow}\ t{\isachardoublequoteclose}\ \isakeyword{and}\ {\isachardoublequoteopen}P\ {\isasymturnstile}\ e\ {\isasymrightarrow}\ s{\isachardoublequoteclose}\isanewline
\ \ \isakeyword{shows}\ {\isachardoublequoteopen}P\ {\isasymturnstile}\ apCon\ e\ Hole\ {\isasymrightarrow}\ t{\isachardoublequoteclose}\isanewline
\isadelimproof
\endisadelimproof
\isatagproof
\isacommand{proof}\isamarkupfalse%
\ {\isacharminus}\isanewline
\ \ \isacommand{from}\isamarkupfalse%
\ assms\ \isacommand{have}\isamarkupfalse%
\ derHole\ {\isacharcolon}\ {\isachardoublequoteopen}P\ {\isasymturnstile}\ s\ {\isasymrightarrow}\ t{\isachardoublequoteclose}\ \isacommand{by}\isamarkupfalse%
\ auto\isanewline
\ \ \isacommand{from}\isamarkupfalse%
\ assms\ \isakeyword{and}\ ctermVals{\isacharbrackleft}of\ {\isachardoublequoteopen}P{\isachardoublequoteclose}\ {\isachardoublequoteopen}e{\isachardoublequoteclose}\ {\isachardoublequoteopen}s{\isachardoublequoteclose}{\isacharbrackright}\ \isacommand{have}\isamarkupfalse%
\ {\isachardoublequoteopen}cterm\ s{\isachardoublequoteclose}\ \isacommand{by}\isamarkupfalse%
\ simp\isanewline
\ \ \isacommand{with}\isamarkupfalse%
\ derHole\ \isakeyword{and}\ lessCTermOrdap\ \isacommand{have}\isamarkupfalse%
\ {\isachardoublequoteopen}t\ {\isasymsqsubseteq}\ s{\isachardoublequoteclose}\ \isacommand{by}\isamarkupfalse%
\ simp\isanewline
\ \ \isacommand{with}\isamarkupfalse%
\ ordapRefl\ crwlPolarity{\isacharbrackleft}of\ {\isachardoublequoteopen}P{\isachardoublequoteclose}\ {\isachardoublequoteopen}e{\isachardoublequoteclose}\ {\isachardoublequoteopen}s{\isachardoublequoteclose}\ {\isachardoublequoteopen}e{\isachardoublequoteclose}\ {\isachardoublequoteopen}t{\isachardoublequoteclose}{\isacharbrackright}\ \isakeyword{and}\ assms\ \isacommand{have}\isamarkupfalse%
\ {\isachardoublequoteopen}P\ {\isasymturnstile}\ e\ {\isasymrightarrow}\ t{\isachardoublequoteclose}\ \isacommand{by}\isamarkupfalse%
\ simp\isanewline
\ \ \isacommand{with}\isamarkupfalse%
\ assms\ \isacommand{show}\isamarkupfalse%
\ {\isacharquery}thesis\ \isacommand{by}\isamarkupfalse%
\ simp\isanewline
\isacommand{qed}\isamarkupfalse%
\endisatagproof
{\isafoldproof}%
\end{isacode}
}{.}
\longText{

\noindent The only new result we have used in that proof is Lemma \isaelem{lessCTermOrdap}, that relates reduction of c-terms with its relation in the approximation order.
\begin{isacode}
\isacommand{lemma}\isamarkupfalse%
\ lessCTermOrdap\ {\isacharcolon}\isanewline
\ \ \isakeyword{assumes}\ {\isachardoublequoteopen}cterm\ t{\isachardoublequoteclose}\isanewline
\ \ \isakeyword{shows}\ {\isachardoublequoteopen}{\isacharparenleft}P\ {\isasymturnstile}\ t\ {\isasymrightarrow}\ s{\isacharparenright}\ {\isacharequal}\ {\isacharparenleft}s\ {\isasymsqsubseteq}\ t{\isacharparenright}{\isachardoublequoteclose}\isanewline
\isadelimproof
\endisadelimproof
\isatagproof
\isacommand{using}\isamarkupfalse%
\ assms\isanewline
\isacommand{by}\isamarkupfalse%
\ {\isacharparenleft}auto\ simp\ add{\isacharcolon}\ lessCTermOrdapL\ lessCTermOrdapR{\isacharparenright}%
\endisatagproof
{\isafoldproof}%
\end{isacode}

To prove it we use the left-to-right of Lemma \isaelem{lessCTermOrdapL} at the beginning of this section, and the right-to-left implication of Lemma \isaelem{lessCTermOrdapR}, which can be easily proved by using the polarity of \crwl.

\begin{isacode}
\isacommand{lemma}\isamarkupfalse%
\ lessCTermOrdapR\ {\isacharcolon}\isanewline
\ \ \isakeyword{assumes}\ {\isachardoublequoteopen}s\ {\isasymsqsubseteq}\ t{\isachardoublequoteclose}\ \isakeyword{and}\ {\isachardoublequoteopen}cterm\ t{\isachardoublequoteclose}\isanewline
\ \ \isakeyword{shows}\ {\isachardoublequoteopen}P\ {\isasymturnstile}\ t\ {\isasymrightarrow}\ s{\isachardoublequoteclose}\isanewline
\isadelimproof
\endisadelimproof
\isatagproof
\isacommand{using}\isamarkupfalse%
\ assms\isanewline
\isacommand{proof}\isamarkupfalse%
\ {\isacharminus}\isanewline
\ \ \isacommand{assume}\isamarkupfalse%
\ less\ {\isacharcolon}\ {\isachardoublequoteopen}s\ {\isasymsqsubseteq}\ t{\isachardoublequoteclose}\isanewline
\ \ \isacommand{from}\isamarkupfalse%
\ assms\ \isakeyword{and}\ ctermRefl\ \isacommand{have}\isamarkupfalse%
\ {\isachardoublequoteopen}P\ {\isasymturnstile}\ t\ {\isasymrightarrow}\ t{\isachardoublequoteclose}\ \isacommand{by}\isamarkupfalse%
\ simp\isanewline
\ \ \isacommand{with}\isamarkupfalse%
\ less\ ordapRefl\ \isakeyword{and}\ crwlPolarity{\isacharbrackleft}of\ {\isachardoublequoteopen}P{\isachardoublequoteclose}\ {\isachardoublequoteopen}t{\isachardoublequoteclose}\ {\isachardoublequoteopen}t{\isachardoublequoteclose}\ {\isachardoublequoteopen}t{\isachardoublequoteclose}\ {\isachardoublequoteopen}s{\isachardoublequoteclose}{\isacharbrackright}\ \isacommand{show}\isamarkupfalse%
\ {\isacharquery}thesis\ \isacommand{by}\isamarkupfalse%
\ simp\isanewline
\isacommand{qed}\isamarkupfalse%
\endisatagproof
{\isafoldproof}%
\end{isacode}}{}
\longText{

\noindent Finally, with lemmas \isaelem{compCRWLHole{\isadigit{1}}} and \isaelem{compCRWLHole{\isadigit{2}}} at hand the proof 
is straightforward.}{}
\longText{

\begin{isacode}
\isacommand{using}\isamarkupfalse%
\ assms\ \isanewline
\isacommand{proof}\isamarkupfalse%
\ {\isacharminus}\isanewline
\ \ \isacommand{have}\isamarkupfalse%
\ {\isachardoublequoteopen}den\ P\ {\isacharparenleft}apCon\ e\ xC{\isacharparenright}\ {\isasymsubseteq}\ {\isacharparenleft}{\isasymUnion}t{\isasymin}den\ P\ e{\isachardot}\ den\ P\ {\isacharparenleft}apCon\ t\ xC{\isacharparenright}{\isacharparenright}{\isachardoublequoteclose}\isanewline
\ \ \isacommand{using}\isamarkupfalse%
\ assms\isanewline
\ \ \isacommand{proof}\isamarkupfalse%
\ {\isacharminus}\isanewline
\ \ \ \ \isacommand{have}\isamarkupfalse%
\ {\isachardoublequoteopen}{\isasymforall}\ t\ {\isasymin}\ den\ P\ {\isacharparenleft}apCon\ e\ xC{\isacharparenright}{\isachardot}\ t\ {\isasymin}\ {\isacharparenleft}{\isasymUnion}s{\isasymin}den\ P\ e{\isachardot}\ den\ P\ {\isacharparenleft}apCon\ s\ xC{\isacharparenright}{\isacharparenright}{\isachardoublequoteclose}\isanewline
\ \ \ \ \isacommand{proof}\isamarkupfalse%
\ {\isacharminus}\isanewline
\ \ \ \ \ \isacommand{{\isacharbraceleft}}\isamarkupfalse%
\isanewline
\ \ \ \ \ \ \isacommand{fix}\isamarkupfalse%
\ t\isanewline
\ \ \ \ \ \ \isacommand{assume}\isamarkupfalse%
\ {\isachardoublequoteopen}t\ {\isasymin}\ den\ P\ {\isacharparenleft}apCon\ e\ xC{\isacharparenright}{\isachardoublequoteclose}\isanewline
\ \ \ \ \ \ \isacommand{hence}\isamarkupfalse%
\ {\isachardoublequoteopen}P\ {\isasymturnstile}\ apCon\ e\ xC\ {\isasymrightarrow}\ t{\isachardoublequoteclose}\ \isacommand{unfolding}\isamarkupfalse%
\ den{\isacharunderscore}def\ \isacommand{by}\isamarkupfalse%
\ simp\isanewline
\ \ \ \ \ \ \isacommand{with}\isamarkupfalse%
\ compCRWL{\isadigit{1}}\ \isacommand{have}\isamarkupfalse%
\ {\isachardoublequoteopen}{\isasymexists}\ s{\isachardot}\ P\ \ {\isasymturnstile}\ e\ {\isasymrightarrow}\ s\ {\isasymand}\ P\ {\isasymturnstile}\ {\isacharparenleft}apCon\ s\ xC{\isacharparenright}\ {\isasymrightarrow}\ t{\isachardoublequoteclose}\ \isacommand{using}\isamarkupfalse%
\ assms\ \isacommand{by}\isamarkupfalse%
\ simp\isanewline
\ \ \ \ \ \ \isacommand{then}\isamarkupfalse%
\ \isacommand{obtain}\isamarkupfalse%
\ s\ \isakeyword{where}\ {\isachardoublequoteopen}P\ \ {\isasymturnstile}\ e\ {\isasymrightarrow}\ s\ {\isasymand}\ P\ {\isasymturnstile}\ {\isacharparenleft}apCon\ s\ xC{\isacharparenright}\ {\isasymrightarrow}\ t{\isachardoublequoteclose}\ \isacommand{by}\isamarkupfalse%
\ auto\isanewline
\ \ \ \ \ \ \isacommand{hence}\isamarkupfalse%
\ {\isachardoublequoteopen}t\ {\isasymin}\ {\isacharparenleft}{\isasymUnion}s{\isasymin}den\ P\ e{\isachardot}\ den\ P\ {\isacharparenleft}apCon\ s\ xC{\isacharparenright}{\isacharparenright}{\isachardoublequoteclose}\ \isacommand{unfolding}\isamarkupfalse%
\ den{\isacharunderscore}def\ \isacommand{by}\isamarkupfalse%
\ auto\isanewline
\ \ \ \ \ \isacommand{{\isacharbraceright}}\isamarkupfalse%
\isanewline
\ \ \ \ \ \ \isacommand{thus}\isamarkupfalse%
\ {\isacharquery}thesis\ \isacommand{by}\isamarkupfalse%
\ auto\isanewline
\ \ \ \ \isacommand{qed}\isamarkupfalse%
\isanewline
\ \ \ \ \isacommand{with}\isamarkupfalse%
\ subset{\isacharunderscore}eq{\isacharbrackleft}of\ {\isachardoublequoteopen}den\ P\ {\isacharparenleft}apCon\ e\ xC{\isacharparenright}{\isachardoublequoteclose}\ {\isachardoublequoteopen}{\isasymUnion}t{\isasymin}den\ P\ e{\isachardot}\ den\ P\ {\isacharparenleft}apCon\ t\ xC{\isacharparenright}{\isachardoublequoteclose}{\isacharbrackright}\ \isacommand{show}\isamarkupfalse%
\ {\isacharquery}thesis\ \isacommand{by}\isamarkupfalse%
\ simp\isanewline
\ \ \isacommand{qed}\isamarkupfalse%
\isanewline
\ \isacommand{moreover}\isamarkupfalse%
\isanewline
\ \ \isacommand{have}\isamarkupfalse%
\ {\isachardoublequoteopen}den\ P\ {\isacharparenleft}apCon\ e\ xC{\isacharparenright}\ {\isasymsupseteq}\ {\isacharparenleft}{\isasymUnion}t{\isasymin}den\ P\ e{\isachardot}\ den\ P\ {\isacharparenleft}apCon\ t\ xC{\isacharparenright}{\isacharparenright}{\isachardoublequoteclose}\isanewline
\ \ \isacommand{using}\isamarkupfalse%
\ assms\isanewline
\ \ \isacommand{proof}\isamarkupfalse%
\ {\isacharminus}\isanewline
\ \ \ \ \isacommand{have}\isamarkupfalse%
\ {\isachardoublequoteopen}{\isasymforall}\ t\ {\isasymin}\ {\isacharparenleft}{\isasymUnion}s{\isasymin}den\ P\ e{\isachardot}\ den\ P\ {\isacharparenleft}apCon\ s\ xC{\isacharparenright}{\isacharparenright}{\isachardot}\ t\ {\isasymin}\ den\ P\ {\isacharparenleft}apCon\ e\ xC{\isacharparenright}{\isachardoublequoteclose}\isanewline
\ \ \ \ \isacommand{proof}\isamarkupfalse%
\ {\isacharminus}\isanewline
\ \ \ \ \ \isacommand{{\isacharbraceleft}}\isamarkupfalse%
\isanewline
\ \ \ \ \ \ \isacommand{fix}\isamarkupfalse%
\ t\isanewline
\ \ \ \ \ \ \isacommand{assume}\isamarkupfalse%
\ \ {\isachardoublequoteopen}t\ {\isasymin}\ {\isacharparenleft}{\isasymUnion}s{\isasymin}den\ P\ e{\isachardot}\ den\ P\ {\isacharparenleft}apCon\ s\ xC{\isacharparenright}{\isacharparenright}{\isachardoublequoteclose}\isanewline
\ \ \ \ \ \ \isacommand{hence}\isamarkupfalse%
\ {\isachardoublequoteopen}{\isasymexists}\ s\ {\isasymin}\ den\ P\ e{\isachardot}\ {\isacharparenleft}t\ {\isasymin}\ den\ P\ {\isacharparenleft}apCon\ s\ xC{\isacharparenright}{\isacharparenright}{\isachardoublequoteclose}\ \isacommand{by}\isamarkupfalse%
\ simp\isanewline
\ \ \ \ \ \ \isacommand{then}\isamarkupfalse%
\ \isacommand{obtain}\isamarkupfalse%
\ s\ \isakeyword{where}\ {\isachardoublequoteopen}{\isacharparenleft}s\ {\isasymin}\ den\ P\ e{\isacharparenright}\ {\isasymand}\ {\isacharparenleft}t\ {\isasymin}\ den\ P\ {\isacharparenleft}apCon\ s\ xC{\isacharparenright}{\isacharparenright}{\isachardoublequoteclose}\ \isacommand{by}\isamarkupfalse%
\ auto\isanewline
\ \ \ \ \ \ \isacommand{hence}\isamarkupfalse%
\ {\isachardoublequoteopen}P\ {\isasymturnstile}\ e\ {\isasymrightarrow}\ s\ {\isasymand}\ P\ {\isasymturnstile}\ {\isacharparenleft}apCon\ s\ xC{\isacharparenright}\ {\isasymrightarrow}\ t{\isachardoublequoteclose}\ \isacommand{unfolding}\isamarkupfalse%
\ den{\isacharunderscore}def\ \isacommand{by}\isamarkupfalse%
\ simp\isanewline
\ \ \ \ \ \ \isacommand{with}\isamarkupfalse%
\ compCRWL{\isadigit{2}}{\isacharbrackleft}of\ {\isachardoublequoteopen}P{\isachardoublequoteclose}\ {\isacharunderscore}\ {\isachardoublequoteopen}t{\isachardoublequoteclose}\ {\isachardoublequoteopen}s{\isachardoublequoteclose}\ {\isachardoublequoteopen}xC{\isachardoublequoteclose}\ {\isachardoublequoteopen}e{\isachardoublequoteclose}{\isacharbrackright}\ \isacommand{have}\isamarkupfalse%
\ {\isachardoublequoteopen}P\ {\isasymturnstile}\ {\isacharparenleft}apCon\ e\ xC{\isacharparenright}\ {\isasymrightarrow}\ t{\isachardoublequoteclose}\ \isacommand{using}\isamarkupfalse%
\ assms\ \isacommand{by}\isamarkupfalse%
\ simp\isanewline
\ \ \ \ \ \ \isacommand{hence}\isamarkupfalse%
\ {\isachardoublequoteopen}t\ {\isasymin}\ den\ P\ {\isacharparenleft}apCon\ e\ xC{\isacharparenright}{\isachardoublequoteclose}\ \isacommand{unfolding}\isamarkupfalse%
\ den{\isacharunderscore}def\ \isacommand{by}\isamarkupfalse%
\ simp\isanewline
\ \ \ \ \ \isacommand{{\isacharbraceright}}\isamarkupfalse%
\isanewline
\ \ \ \ \ \ \isacommand{thus}\isamarkupfalse%
\ {\isacharquery}thesis\ \isacommand{by}\isamarkupfalse%
\ auto\isanewline
\ \ \ \ \isacommand{qed}\isamarkupfalse%
\isanewline
\ \ \ \ \isacommand{with}\isamarkupfalse%
\ subset{\isacharunderscore}eq{\isacharbrackleft}of\ {\isachardoublequoteopen}{\isasymUnion}t{\isasymin}den\ P\ e{\isachardot}\ den\ P\ {\isacharparenleft}apCon\ t\ xC{\isacharparenright}{\isachardoublequoteclose}\ {\isachardoublequoteopen}den\ P\ {\isacharparenleft}apCon\ e\ xC{\isacharparenright}{\isachardoublequoteclose}{\isacharbrackright}\ \isacommand{show}\isamarkupfalse%
\ {\isacharquery}thesis\ \isacommand{by}\isamarkupfalse%
\ simp\isanewline
\ \ \isacommand{qed}\isamarkupfalse%
\isanewline
\isacommand{ultimately}\isamarkupfalse%
\isanewline
\ \ \isacommand{show}\isamarkupfalse%
\ {\isacharquery}thesis\ \isacommand{using}\isamarkupfalse%
\ subset{\isacharunderscore}antisym{\isacharbrackleft}of\ {\isachardoublequoteopen}den\ P\ {\isacharparenleft}apCon\ e\ xC{\isacharparenright}{\isachardoublequoteclose}\ {\isachardoublequoteopen}{\isasymUnion}t{\isasymin}den\ P\ e{\isachardot}\ den\ P\ {\isacharparenleft}apCon\ t\ xC{\isacharparenright}{\isachardoublequoteclose}{\isacharbrackright}\ \isacommand{by}\isamarkupfalse%
\ simp\isanewline
\isacommand{qed}\isamarkupfalse%
\endisatagproof
{\isafoldproof}%
\end{isacode}}{

}
%
Again, while theorem~\isaelem{compCRWL} requires the context to have just
one hole, it does not assume the linearity or constructor discipline
of the program. This came as a surprise to us, and initially made us
doubt about the accuracy of our formalization of \crwl. But it turns
out that although \crwl\ is designed to work with \crwl-programs, that
fulfil these restrictions, it can also be applied to general programs.
For those programs some properties, such as the theorems
\isaelem{crwlClosedCSubst}, \isaelem{crwlPolarity}, and
\isaelem{compCRWL} still hold, but other fundamental properties do
not, in particular the strong adequacy results w.r.t.\ its operational
counterparts of~\cite{GHLR99,ppdp2007,AHHOV05}.
The point is that for those programs \crwl\ does not deliver the
``intended semantics'' anymore. And this is not strange, because that semantics was intended
with \crwl-programs in mind.
For example, consider the non linear
program $\prog = \{f(X,X) \tor a\}$. 
There is a \crwl-proof for the statement $\prog \vdash f(a,b) \clto
a$ 
but this value cannot be computed in any of the operational notions
of~\cite{GHLR99,ppdp2007,AHHOV05} nor in any implementation of FLP, in
which the independence of the matching process of the arguments --- derived from left-linearity of program rules --- is
assumed. It is also not very natural that $f(a,b)$ could yield the value $a$ for the arguments $a$ and $b$ being different values, which implies that the semantics defined by \crwl\ for non left-linear programs is pretty odd. But that is not a big problem, because we only care about the properties of \crwl\ for the kind of programs it has been designed to work with. 
And if it enjoys some interesting properties for a bigger class of programs that is fine, because that nice properties will be inherited by the class of \crwl-programs.

On the other hand,
for programs not following the constructor discipline, we will not even be
%
%
able to have a matching for an argument of a rule which is not a
constructor, because in the rule \crule{OR} we have to reduce every
argument of a function call to a value, which will be a c-term by
Lemma \isaelem{ctermVals}\longText{}{ (see the extended version of this paper)}, and so could never be an instance of
expression containing function symbols. Thus, the rule \crule{OR} could not be used for program rules not following the constructor discipline.
\longText{\\

We will end this section with the following result about the replaceability of expressions in one-hole contexts, and easy consequence of the compositionality of \crwl.
\begin{isacode}
\isacommand{theorem}\isamarkupfalse%
\ cntxtReplace\ {\isacharcolon}\ {\isachardoublequoteopen}{\isacharparenleft}den\ P\ e{\isadigit{1}}\ {\isacharequal}\ den\ P\ e{\isadigit{2}}{\isacharparenright}\ {\isacharequal}\ \ {\isacharparenleft}{\isasymforall}\ xC{\isachardot}\ oneHole\ xC\ {\isacharminus}{\isacharminus}{\isachargreater}\ {\isacharparenleft}den\ P\ {\isacharparenleft}apCon\ e{\isadigit{1}}\ xC{\isacharparenright}\ {\isacharequal}\ den\ P\ {\isacharparenleft}apCon\ e{\isadigit{2}}\ xC{\isacharparenright}{\isacharparenright}{\isacharparenright}{\isachardoublequoteclose}\isanewline
\isadelimproof
\endisadelimproof
\isatagproof
\isacommand{proof}\isamarkupfalse%
\ {\isacharminus}\isanewline
\ \ \isacommand{{\isacharbraceleft}}\isamarkupfalse%
\isanewline
\ \ \isacommand{have}\isamarkupfalse%
\ {\isachardoublequoteopen}{\isacharparenleft}den\ P\ e{\isadigit{1}}\ {\isacharequal}\ den\ P\ e{\isadigit{2}}{\isacharparenright}\ {\isacharminus}{\isacharminus}{\isachargreater}\ {\isacharparenleft}{\isasymforall}\ xC{\isachardot}\ oneHole\ xC\ {\isacharminus}{\isacharminus}{\isachargreater}\ {\isacharparenleft}den\ P\ {\isacharparenleft}apCon\ e{\isadigit{1}}\ xC{\isacharparenright}\ {\isacharequal}\ den\ P\ {\isacharparenleft}apCon\ e{\isadigit{2}}\ xC{\isacharparenright}{\isacharparenright}{\isacharparenright}{\isachardoublequoteclose}\isanewline
\ \ \isacommand{by}\isamarkupfalse%
\ {\isacharparenleft}auto\ simp\ add{\isacharcolon}\ compCRWL{\isacharparenright}\isanewline
\ \ \isacommand{{\isacharbraceright}}\isamarkupfalse%
\isanewline
\ \isacommand{moreover}\isamarkupfalse%
\isanewline
\ \ \isacommand{{\isacharbraceleft}}\isamarkupfalse%
\isanewline
\ \ \isacommand{have}\isamarkupfalse%
\ {\isachardoublequoteopen}{\isacharparenleft}{\isasymforall}\ xC{\isachardot}\ \ oneHole\ xC\ {\isacharminus}{\isacharminus}{\isachargreater}\ {\isacharparenleft}den\ P\ {\isacharparenleft}apCon\ e{\isadigit{1}}\ xC{\isacharparenright}\ {\isacharequal}\ den\ P\ {\isacharparenleft}apCon\ e{\isadigit{2}}\ xC{\isacharparenright}{\isacharparenright}{\isacharparenright}\ {\isacharminus}{\isacharminus}{\isachargreater}\ {\isacharparenleft}den\ P\ e{\isadigit{1}}\ {\isacharequal}\ den\ P\ e{\isadigit{2}}{\isacharparenright}{\isachardoublequoteclose}\isanewline
\ \ \isacommand{proof}\isamarkupfalse%
\isanewline
\ \ \ \ \isacommand{assume}\isamarkupfalse%
\ h{\isadigit{1}}\ {\isacharcolon}\ {\isachardoublequoteopen}{\isasymforall}\ xC{\isachardot}\ \ oneHole\ xC\ {\isacharminus}{\isacharminus}{\isachargreater}\ {\isacharparenleft}den\ P\ {\isacharparenleft}apCon\ e{\isadigit{1}}\ xC{\isacharparenright}\ {\isacharequal}\ den\ P\ {\isacharparenleft}apCon\ e{\isadigit{2}}\ xC{\isacharparenright}{\isacharparenright}{\isachardoublequoteclose}\isanewline
\ \ \ \ \isacommand{hence}\isamarkupfalse%
\ {\isachardoublequoteopen}den\ P\ {\isacharparenleft}apCon\ e{\isadigit{1}}\ Hole{\isacharparenright}\ {\isacharequal}\ den\ P\ {\isacharparenleft}apCon\ e{\isadigit{2}}\ Hole{\isacharparenright}{\isachardoublequoteclose}\isanewline
\ \ \ \ \isacommand{proof}\isamarkupfalse%
\ {\isacharminus}\isanewline
\ \ \ \ \ \ \isacommand{have}\isamarkupfalse%
\ {\isachardoublequoteopen}oneHole\ Hole{\isachardoublequoteclose}\ \isacommand{unfolding}\isamarkupfalse%
\ oneHole{\isacharunderscore}def\ \isacommand{by}\isamarkupfalse%
\ simp\isanewline
\ \ \ \ \ \ \isacommand{with}\isamarkupfalse%
\ h{\isadigit{1}}\ \isacommand{show}\isamarkupfalse%
\ {\isacharquery}thesis\ \isacommand{by}\isamarkupfalse%
\ blast\isanewline
\ \ \ \ \isacommand{qed}\isamarkupfalse%
\isanewline
\ \ \ \ \isacommand{thus}\isamarkupfalse%
\ {\isachardoublequoteopen}den\ P\ e{\isadigit{1}}\ {\isacharequal}\ den\ P\ e{\isadigit{2}}{\isachardoublequoteclose}\ \isacommand{by}\isamarkupfalse%
\ simp\isanewline
\ \ \isacommand{qed}\isamarkupfalse%
\isanewline
\ \ \isacommand{{\isacharbraceright}}\isamarkupfalse%
\isanewline
\ \isacommand{ultimately}\isamarkupfalse%
\ \isacommand{show}\isamarkupfalse%
\ {\isacharquery}thesis\ \isacommand{by}\isamarkupfalse%
\ auto\isanewline
\isacommand{qed}\isamarkupfalse%
\endisatagproof
{\isafoldproof}%
\end{isacode}}{}



\section{Conclusions}\label{conclusions}
This paper presented a formalization of the essentials of
\crwl~\cite{GHLR96,GHLR99}, a well-known semantic framework for
functional logic programming, in the interactive proof assistant
Isabelle/HOL. We chose that particular logical framework for its
stability and its extensive libraries. The Isar proof language allowed
us to structure the proofs so that they become quite elegant and
readable, as can be observed by looking at the Isabelle code.

Our formalization is generic with respect to syntax,
\longText{%
  in the sense that a previously given signature and program is not
  assumed,%
}{}
and includes important auxiliary notions like substitutions or
contexts. This is in contrast to previous
work~\cite{ClevaLL04,ClevaPita06}
\longText{%
  where some work about formalization of CRWL was reported, but
  focused on formalizing the semantics of each concrete program, as a
  way of proving concrete program properties.
}{%
  that focused on formalizing the semantics of each concrete program.%
}
In contrast, our paper focuses on developing the metatheory of the
formalism, allowing us to obtain results that are more general and
also more powerful: we formally prove essential properties of the
paradigm like \emph{polarity} or \emph{compositionality} of the
\crwl-semantics.
\longText{%
  Of course, such general properties hold for each concrete program,
  but nothing similar was achieved in the above mentioned previous
  works.%
}{}
We plan to extend our theories so that we will be able to reason about
properties of concrete programs by deriving theorems that express
verification conditions in the line of those stated
in~\cite{ClevaLL04,ClevaPita06}.


While developing the formalization we realized an interesting fact not
pointed out before: properties like polarity or compositionality do
not depend on the constructor discipline and left-linearity imposed to
programs. However, such requirements will certainly play an essential
role when extending our work to formally relate the \crwl-semantics
with operational semantics like the one developed in~\cite{ppdp2007},
one of our intended subjects of future work. We think that could be
interesting in several ways. First of all it would be a further step
in the direction of challenge~3
of~\cite{DBLP:conf/tphol/AydemirBFFPSVWWZ05}, ``Testing and Animating
wrt the Semantics'', because we would end up getting an interpreter of
\crwl\ during the process. We should then also formalize the
evaluation strategy for the operational semantics, obtaining an
Isabelle proof of its optimality. Finally there are
precedents~\cite{LRSflops08,ppdp2007} of how the combination of a
denotational and operational perspective is useful for general
semantic reasoning in FLP.
%
%
%



\end{document}